\documentclass[twocolumn,useAMS,usenatbib,a4paper]{mn2e}
\usepackage{xspace}
\usepackage{hyperref}
\usepackage{aas_macros}
\usepackage{url}
\usepackage{amsmath}
\usepackage{amssymb}
\usepackage[noabbrev, capitalise]{cleveref}
\usepackage{graphicx}
\usepackage{xcolor}
\usepackage{subcaption}
\usepackage{multirow}
\usepackage{booktabs}
\usepackage{appendix}

\newcommand{\E}{\times10}
\newcommand{\jbca}{{Jodrell Bank Centre for Astrophysics, School of Physics and Astronomy, The University of Manchester, Manchester M13 9PL, UK}}
\newcommand{\clean}{\textsc{clean}\xspace}
\crefname{appsec}{Appendix}{Appendices}
\usepackage{float}
\restylefloat{table}

\title[COSMOS Radio-Optical Galaxy Shapes]{Radio-Optical Galaxy Shape Correlations in the\\COSMOS Field} 
\author[Tunbridge \emph{et al.}]{Ben Tunbridge\thanks{E-mail: benjamin.tunbridge@manchester.ac.uk}, Ian Harrison, Michael L. Brown\\\jbca}

\begin{document}
\maketitle

\begin{abstract}


\noindent We investigate the correlations in galaxy shapes between optical and radio wavelengths using archival observations of the COSMOS field. Cross-correlation studies between different wavebands will become increasingly important for precision cosmology as future large surveys may be dominated by systematic rather than statistical errors. In the case of weak lensing, galaxy shapes must be measured to extraordinary accuracy (shear systematics of $< 0.01\%$) in order to achieve good constraints on dark energy parameters. By using shape information from overlapping surveys in optical and radio bands, robustness to systematics may be significantly improved without loss of constraining power. Here we use HST-ACS optical data, VLA radio data, and extensive simulations to investigate both our ability to make precision measurements of source shapes from realistic radio data, and to constrain the intrinsic astrophysical scatter between the shapes of galaxies as measured in the optical and radio wavebands. By producing a new image from the VLA-COSMOS L-band radio visibility data that is well suited to galaxy shape measurements, we are able to extract precise measurements of galaxy position angles. Comparing to corresponding measurements from the HST optical image, we set a lower limit on the intrinsic astrophysical scatter in position angles, between the optical and radio bands, of $\sigma_\alpha > 0.212\pi$ radians (or $38.2^{\circ}$) at a $95\%$ confidence level.

\end{abstract}

\section{Introduction}\label{Introduction}
Weak gravitational lensing analyses exploit the coherent distortion of galaxy shapes induced by the gravitational potential along the line of sight to constrain the distribution and evolution of massive structures in the Universe, providing an excellent probe of cosmology \citep[see e.g.][for a review]{2015RPPh...78h6901K}. Conducting precision cosmology tests with weak lensing (e.g.~selecting between competing models of dark energy) requires high number densities ($n_{\rm gal} \gtrsim 1 \, \mathrm{arcmin}^{-2}$) of high redshift ($z \gtrsim 1$) sources and analysis tools to measure their shapes to extraordinary accuracy. These attributes have been available to surveys at optical wavelengths for a number of years, with useful cosmological results beginning to emerge from CFHTLens \citep{2016MNRAS.456.1508K}, DES-SV \citep{2015arXiv150705552T} and DLS \citep{2015arXiv151003962J}. As source number densities increase in these and future experiments (and statistical uncertainties decrease) it is the systematic uncertainties which will come to dominate. One way of tackling the problem of such systematics is through multi-wavelength investigations, which have the potential to address both instrumental and astrophysical contamination. In particular, telescopes operating at radio wavelengths are about to undergo a significant leap in survey speed and sensitivity and, ultimately, with the Square Kilometre Array (SKA)\footnote{\url{http://www.skatelescope.org}} will be capable of world-leading weak lensing cosmology alone \citep{2015aska.confE..23B, 2016arXiv160103948B}. In addition, cross-correlation weak lensing studies between radio and optical wavebands appear highly promising, providing comparable cosmological constraints as traditional single-waveband experiments, but with the significant advantage of being much more robust to wavelength-dependent systematic effects \citep{2016MNRAS.456.3100D, 2016arXiv160103947H}.

In this paper we focus on a particular aspect of these cross-correlation studies: the correlation between the optical and radio shapes of those galaxies which will be common to both catalogues. As discussed in \cite{2016arXiv160103947H} this shape co-variance constitutes a noise term on the measured weak lensing shear power spectra, and needs to be well understood to form robust cosmological parameter constraints. Furthermore, literature results on these shape correlations are divided, with \cite{2009MNRAS.399.1888B} finding a significant correlation in orientations between the shapes of objects detected in the SDSS and FIRST catalogues, and \cite{2010MNRAS.401.2572P} finding almost no correlation in the optical and radio shapes of galaxies in the Hubble Deep Field North (HDF-N). Here we measure and compare the optical and radio shapes of sources in the COSMOS field, which has deep data available in both wavebands. To facilitate an accurate comparison, we have implemented a significant re-analysis of the radio VLA-COSMOS data set. Our re-analysis significantly reduces the level of galaxy shape systematics which we found to be present in the previous analysis of \cite{2007ApJS..172...46S} (though we note that their analysis was primarily focused on measuring faint number counts and was not optimised for extracting galaxy shapes). 

When conducting any study which relies upon accurate galaxy shape measurements, it is important to appreciate that the data generated in radio observations is fundamentally different to that generated by imaging telescopes in optical wavebands. Radio telescopes are typically operated as interferometers, consisting of many individual antennas, from which the signals are correlated to form `visibilities'. For a pair of antennas a projected distance $d$ apart, the recorded visibility is a measurement of the flux at a specific spatial frequency on the sky. For interferometers with $N$ antennas, $N(N-1)/2$ correlations can be formed between the different baseline lengths, sampling a large number of scales and allowing us to reconstruct a partially sampled Fourier transform of the sky brightness distribution \citep[when the assumptions of the Van Cittert–-Zernike theorem,][are satisfied]{2009MNRAS.395.1558C}. This can provide distinct advantages, allowing for very good angular resolution (set by the longest baseline in the array) with a comparably large maximum field of view (set by the primary beam of the individual antennas).

The incomplete sampling of spatial scales means the data must be further processed to obtain an estimate of the corresponding image plane information. The resultant Point Spread Function (PSF) from a radio interferometer (commonly referred to as a `dirty beam') is highly deterministic at $\gtrsim 700 \, \mathrm{MHz}$ frequencies (as it is set by the known sampling of the Fourier plane -- a useful property for weak lensing) but can have significant structure, with appreciable sidelobes extending across the entire sky. The convolution of this complicated beam with an unknown sky creates the `dirty image' (the simple Fourier transform to image space of the data). Deconvolving the dirty beam from the dirty image to enable source identification, flux and shape measurement is then a difficult process.

A commonly used algorithm for this purpose is known as H\"ogbom-\clean. This algorithm assumes that any section of the sky is made up of a finite number of delta function point sources. Beginning with the dirty image, \clean iteratively finds the position and brightness of the brightest source in the field and subtracts flux, corresponding to the convolution of this point source with the dirty beam, directly from the visibility data.
This is repeated across the image until a residual map remains with \clean components, and produces a `\clean' map \citep[see][for a detailed description and motivation]{1974A&AS...15..417H}. Additionally this procedure helps to reduce sidelobe artifacts across the radio map which can be problematic, especially when a large dynamical range in source fluxes is considered.

The \clean algorithm works well at producing plausible radio images and placing sources at reliable positions. However \clean is a non-linear deconvolution technique which assumes that missing information, due to a partially sampled visibility (or Fourier) plane, can be modelled as a set of point sources. This inherent assumption may result in significant systematics when attempting to measure precision morphology as is necessary for weak lensing cosmology. \cite{2015aska.confE..30P} have tested the performance of \clean on SKA-like simulations and have found the performance to be orders of magnitude away from what is necessary for systematic uncertainties from shape measurement to be sub-dominant to statistical ones when measuring cosmic shear. In this paper, we have adopted a \clean-based imaging pipeline in our re-analysis of the VLA-COSMOS data. However, mindful of the above concerns around potential shape biases being induced by the \clean processing, we have also carried out extensive simulations to quantify the level of shape measurement bias we expect in our newly-created image. 

In addition to minimising systematics through cross-correlation, radio weak lensing also presents further potential advantages through additional information from polarisation \citep{2011MNRAS.410.2057B, 2015MNRAS.451..383W} and rotational velocity maps \citep{2006ApJ...650L..21M} which may be used to reduce the shot noise from intrinsic galaxy shapes and also mitigate against intrinsic alignment systematics -- a key astrophysical systematic due to correlations of galaxy shapes being imprinted during their formation process. To date two attempts have been made to measure a weak lensing signal in radio data alone. In the first, \cite{2004ApJ...617..794C} were able to detect a lensing aperture mass signal at a significance of  $3.6\sigma$. In this case the Fourier-plane data from the FIRST radio survey \citep{1995ApJ...450..559B} was used to estimate the shear on radio sources directly, without imaging, modelling them using Fourier-plane shapelets \citep{2002ApJ...570..447C}. The second study, by \cite{2010MNRAS.401.2572P}, used combined radio observations from the VLA and MERLIN interferometers to extract galaxy shapes from radio images constructed by the H\"ogbom-\textsc{clean} method discussed above, where the low absolute number of detected sources precluded a significant weak lensing detection.

This paper is structured as follows. In \cref{The COSMOS Field} we introduce the COSMOS data used in this study and outline the data reduction steps that we have applied. In \cref{Shape Measurements}, we describe our technique for creating simulated radio datasets and we also describe our approach to measuring the radio galaxy shapes, which includes the extraction of a model of the VLA-COSMOS PSF from the simulated datasets. An assessment of the accuracy of the shape parameter recovery from the radio simulations is provided in \cref{simulation_err}. In \cref{vla_shape_measurements} we present the shape measurements from the real VLA-COSMOS data. We then investigate the correlation between the radio-derived and optical-derived position angles of galaxies within the COSMOS field in \cref{Multi-Waveband Shape Comparison}. Finally, in \cref{concl} we summarise our conclusions.

\section{Data}\label{The COSMOS Field}
The COSMOS field is a $2 \, \mathrm{deg}^{2}$ patch of sky which has already been subject to extensive multi-wavelength analysis, with data available from a large number of telescopes including the Hubble Space Telescope (HST, optical), the Very Large Array (VLA, radio), as well as the Spitzer (infrared), GALEX (UV), and XMM/Newton (X-ray) satellites. In this study we focus on the radio and optical observations, taken with the VLA at L-band (1.4~GHz) in A- and C-array configuration, and with the Advanced Camera for Surveys (ACS) instrument onboard the HST, respectively. 

\subsection{VLA radio observations}
The VLA L-band observations consist of a combination of A- and C-array observations sub-divided into two intermediate frequency (IF) chunks which are centred at 1.3649 and 1.4351 GHz, with each of the IFs further divided into 6 channels each averaging over a 3.125 MHz bandwidth. The combination of the A- and C-array observations allows a wider range of spatial scales to be probed. (The accessible spatial scales are dependent upon antenna configuration -- A-array corresponds to smaller scales, and C-array to larger scales in this case). The observation time totals 240 hours and 19.5 hours in A- and C-array configurations respectively. In total 23 pointings were required to cover the COSMOS field, providing a separation of 15$^{\prime}$ between the individual field centres. The VLA observations and accompanying data reduction and imaging was initially conducted by \cite{2007ApJS..172...46S}, where more information on the observation strategy can be found.

\begin{figure}
 \centering
 \includegraphics[width=8cm]{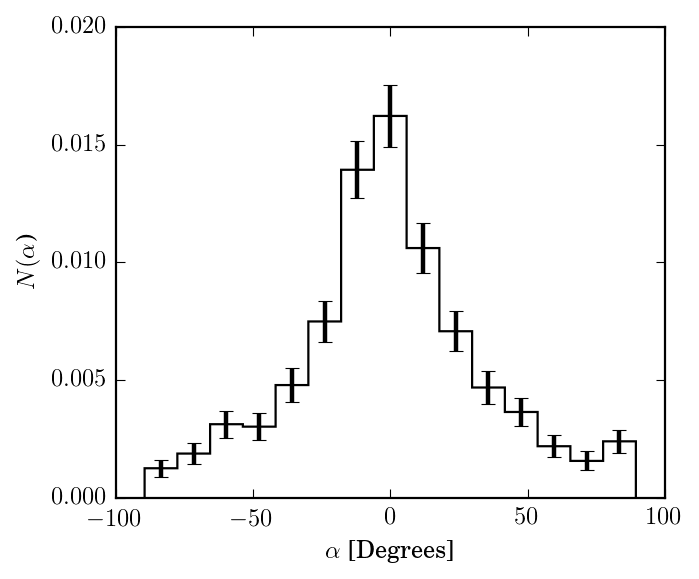}
 \caption[Radio $\alpha$ parameter from original image.]{A histogram of the position angles of all resolved sources detected in the VLA-COSMOS radio image of \cite{2007ApJS..172...46S}. The galaxy position angles ($\alpha$) were measured using the \textsc{im3shape} shape measurement software. The measured distribution is clearly inconsistent with the expected random uniform distribution. Poisson error bars are included.}\label{fig:PA_original}
\end{figure}

We performed an initial examination of the radio image as produced by \cite{2007ApJS..172...46S} in order to assess its suitability for extracting galaxy shapes.\footnote{We performed this assessment on the publicly available image produced by  \cite{2007ApJS..172...46S}, which we downloaded from \url{http://www.mpia.de/COSMOS/}.}. The results of this assessment are illustrated in \cref{fig:PA_original}, which shows a histogram of the position angles ($\alpha$) of all resolved sources detected in the image. To extract the galaxy shapes, we used the \textsc{im3shape} package \citep{2013MNRAS.434.1604Z}, described in further detail in \cref{Shape Measurements}. Note that in this study we define $\alpha$ to be the position angle of the galaxy measured North of East (i.e. counter-clockwise). 

In the absence of lensing and/or intrinsic alignment effects, we expect the galaxies to be randomly orientated and the position angle histogram to be consistent with a uniform distribution. 
Clearly, a large deviation from the expected random uniform distribution is observed. This is likely to be due to systematic effects introduced during the image creation and/or shape measurement steps. 
For example, as discussed in \cref{Introduction}, we expect systematic effects to be introduced at some level due to the use of the \clean-based imaging routine -- the \textsc{aips} task `\textsc{imagr}'\footnote{Wide field imaging algorithm via the H\"ogbom-\clean method.} -- that was used in the creation of the \cite{2007ApJS..172...46S} image. Noting again that the \cite{2007ApJS..172...46S} study was not focused on shape measurements, and motivated by the results shown in \cref{fig:PA_original}, we have performed our own imaging analysis of the VLA-COSMOS visibility data. While our re-analysis also makes use of the \clean algorithm, implementing the data reduction and imaging ourselves enables us to accurately quantify the shape measurement bias due to the imaging pipeline through simulations. 

\subsubsection{Data Reduction}\label{Data Reduction}
For the purposes of this work, we require a precise quantification of any shape systematics caused by the data reduction and imaging processes. This necessitates the use of the simulations detailed in \cref{simulation_err} and the construction of a coherent pipeline which may be run identically on both simulations and real data. For the radio data we begin from partially processed continuous visibility data provided to us by E.~Schinnerer and V.~Smolcic (via private communication with C.~Hales). Although this data was partially reduced, we found it necessary to apply further data reduction steps in order to arrive at a fully reduced and calibrated visibility dataset. 

We have used the Common Astronomy Software Applications (\textsc{casa}; \cite{2007ASPC..376..127M}) packages, following the calibration steps outlined in \cite{1999ASPC..180.....T}, and implemented in \cite{2007ApJS..172...46S}.

We first removed Radio Frequency Interference (RFI) from the data by flagging visibilities which contained RFI at higher levels than the sky signal. After clipping the data to remove all entries with amplitude $>0.9 \, \mathrm{Jy}$ we further visually inspected the data to remove structures corresponding to RFI. The amount of data removed was $9.9\%$ of the initial data. The remaining $(u,v,w)$ locations are then used throughout the rest of our analysis, including for the simulations.

Calibration errors in the visibilities were accounted for by intercalated observations of calibration sources with known fluxes and morphologies, with the calibration solutions interpolated on to the observations of the target field. Three calibration sources were observed: primary flux calibration source 0521+166; primary phase calibrator 1024-008; and, as a secondary calibrator in both flux and phase, 0925+003.

We proceeded through a series of reduction steps from a standard radio astronomy data reduction cookbook as described in, for example \cite{1999ASPC..180...79F}:
\begin{itemize}
\item{Delay (phase) calibration}
\item{Initial bandpass calibration}
\item{Gain calibration of calibration sources}
\item{Flux derivation for calibration sources}
\item{Bandpass calibration with spectral indices}
\item{Update amplitude and phase calibration solutions}
\item{Apply calibration solutions to target fields}
\end{itemize}
The result of these steps was initially calibrated visibility data, ready for image-plane calibration and final imaging, as described in the next section.

\subsubsection{Imaging}\label{Imaging}
Provided with the calibrated visibilities for all 23 VLA pointings we developed an imaging pipeline implemented as a series of \textsc{casa} tasks. Each of the 23 pointings was imaged separately using the \clean task, which deconvolves the complex radio interferometer beam following the H\"ogbom-\clean algorithm introduced in \cref{Introduction}, and described further in \cite{1974A&AS...15..417H}.

Residual sidelobe noise from one bright source (RA 10:02:51.242, DEC +2:42:48.70) was removed from multiple pointings using a technique known as `peeling'. Peeling is a form of direction-dependent calibration which involves using the best available model to subtract all other sources from the data except the off-axis peeling target. The technique then models only the off-axis source (via self-calibration) before subtracting the fitted model from the original data set with all sources included. A detailed description can be found in \cite{2004SPIE.5489..817N}. When imaging the data, we ran the pipeline using two separate weighting schemes in order to evaluate the effectiveness of both. These weighting schemes are used in gridding -- the step which places the continuous visibilities onto a regular grid amenable to Fast Fourier Transforms -- and the choice of a particular scheme can affect the radio map on the arcsec scales to which we are sensitive. `Natural' weighting equally weights all of the continuous visibilities and optimises for point-source sensitivity (but sacrifices resolution), whilst `uniform' weighting inversely weights according to the density at which the UV plane is being sampled, and optimises for resolution at the expense of sensitivity \citep{1999ASPC..180.....T}.


Since each pointing has a relatively large field of view (radius of 20.8$^{\prime}$), curvature across the field of view needs to be accounted for: the Fourier transform performed by the interferometer is no longer in two dimensions ($u,v$) only but includes a third $w$ term. To account for this, we use the `\textsc{wprojplanes}' tool within \clean to automatically assign $w$-projection planes for convolution of non-coplanar effects \citep[see][for a full description of $w$-projection]{2008ISTSP...2..647C}.


For each weighting scheme (natural and uniform), the data was gridded onto pixels corresponding to $0.35^{\prime\prime}$ in the image plane and covering the full primary beam field of view. `\textsc{clean boxes}' were interactively added to areas of likely identified bright sources to assist the imaging algorithm, and this was repeated as more sources became visible in the residual map. A loop gain of $0.05$ is applied during each iteration. This iterative imaging process was regarded as converged either after $100,000$ iterations, or when the RMS of the residual image after model subtraction was $<45\, \mu$Jy.

\begin{figure}
 \centering
 \includegraphics[width=0.5\textwidth,angle=0]{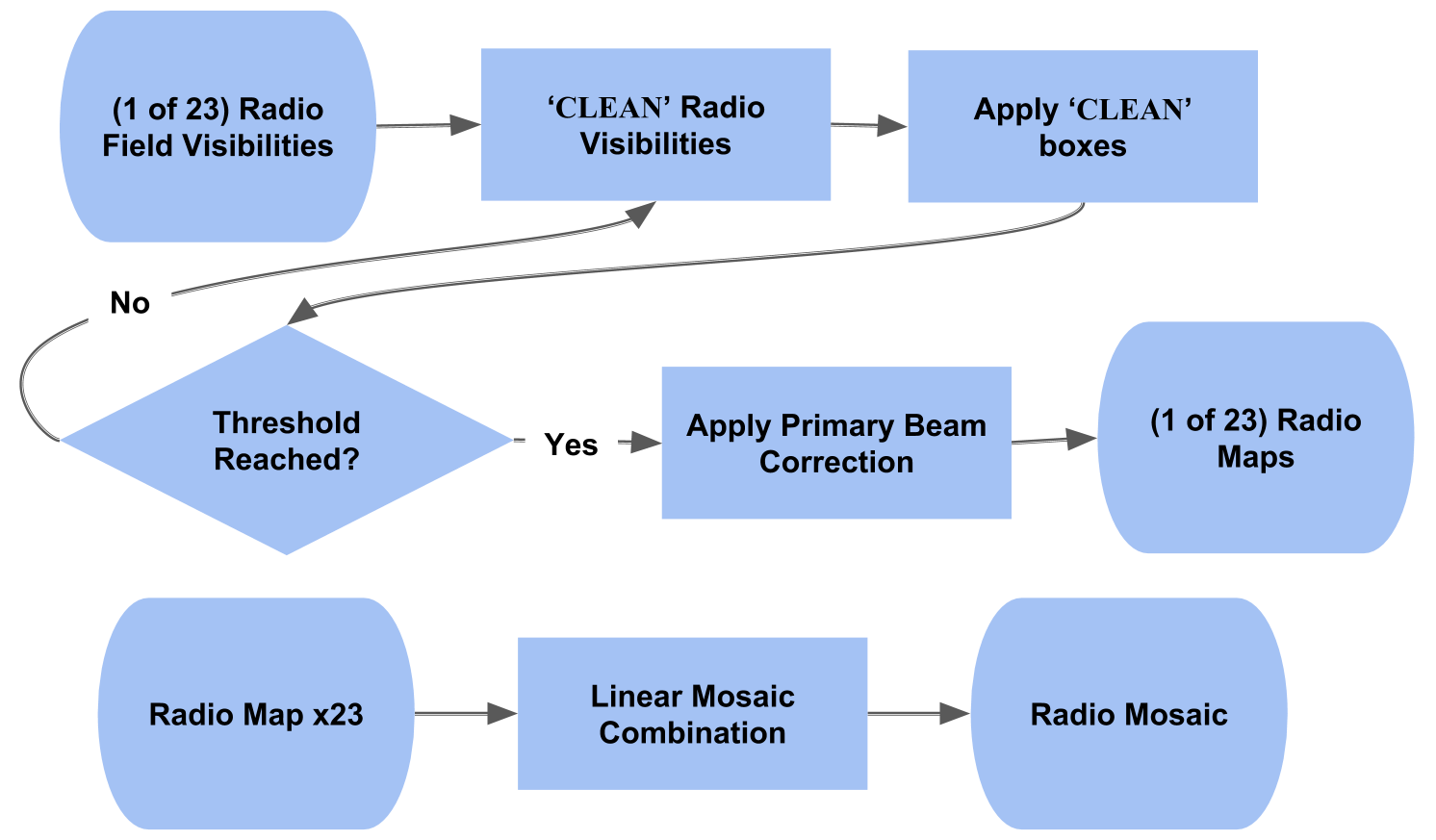}
 \caption[Radio Imaging Pipeline]{A flow chart depicting the radio imaging pipeline. Each of the 23 fields are first run through the top rung. Following this the 23 fields are combined  into a linear mosaic.}
 \label{fig:Radio Imaging Pipeline}
\end{figure}

For each pointing, the output of the above procedure was a map of the entire field of view ($16,000\times16,000$ pixels). To this map, we applied a correction for the primary beam pattern using a gain level of 0.5, which corresponds to a radius of 20.8$^{\prime}$. Finally the 23 individual pointings were combined into a mosaiced image using the \textsc{imager.linearmosaic} tool. Our imaging pipeline is shown schematically in  \cref{fig:Radio Imaging Pipeline} and a full list of \clean inputs can be found in the imaging pipeline script available at \url{https://github.com/bentunbridge/cosmos-radio-optical}.

The image produced using natural weighting has a beam central lobe size with a full width at half maximum (FWHM) of $1.5^{\prime\prime}$ $\times$ $1.4^{\prime\prime}$, at an orientation of $70^{\circ}$, while the image obtained using uniform weighting has a beam central lobe size with FWHM $1.45^{\prime\prime}$ $\times$ $1.3^{\prime\prime}$ at an orientation of $-20^{\circ}$. The position angles here are defined in a counter-clockwise sense, North of East. One could, in principle, use these recovered beam parametrisations as an approximation to the observational PSF during galaxy shape estimation. However as shown in \cref{Simulation} this can be improved upon by directly measuring the PSF from images reconstructed from simulations that include injected unresolved sources (point sources) placed across the field. The RMS noise level in our final mosaiced image was 33 (59) $\mu$Jy/beam and 28 (52) $\mu$Jy/beam in the central $1$ $(2)$ deg$^{2}$ region of the field for the natural (uniform) weighting scheme. Although this is $\sim 3$ times the noise level achieved in the \cite{2007ApJS..172...46S} study we find only a minor drop in resolved source counts ($\sim 6\%$) while, as we demonstrate in \cref{simulation_err}, we find our image is more robust in terms of shape systematics effects. Implementing our own data reduction and imaging pipeline also allows us to accurately model any analysis-induced systematic effects through simulations.  

\subsection{HST-ACS survey data}\label{HST data}
For our optical shape measurements, we use the observations collected with the ACS instrument on board the HST, as described in \cite{2007ApJS..172..196K}. These observations were conducted over a total of 583 HST orbits, covering a field that is 1.64 deg$^{2}$ in area. 

The observations were collected using the F814W filter and reach a limiting magnitude of $\mathrm{I}_{\mathrm{AB}} = 27.2$ mag at $10\sigma$, and the angular resolution is $0.05$ arcsec \citep{2010ApJ...708..202M}. The science data used in this study was produced by \cite{2007ApJS..172..196K} and consists of $575$ pointings across the entire mosaic. Additionally, for each pointing the raw exposures were combined using the MultiDrizzle pipeline \citep{2003hstc.conf..337K} to improve the pixel scale to $0.03$ arcsec, which we use in our analysis. The output MultiDrizzled image cutouts are available at \url{http://irsa.ipac.caltech.edu/Missions/cosmos.html}.

\section{Shape Measurements}\label{Shape Measurements}
\subsection{Modelling the effective radio PSF using simulations}\label{Simulation}
To date, many shape measurement techniques have been developed and optimised for weak lensing surveys, predominantly at optical wavelengths. A key systematic effect to account for in these techniques is the correct deconvolution of the PSF of the observations, and it is equally important to account for such effects in radio observations. As discussed in \cref{Introduction}, the PSF applied to the sky (the `dirty beam') by a radio interferometer is precisely known and highly deterministic (and mostly unaffected by stochastic variation in the atmosphere). However, this is only the correct PSF for the dirty map and has significant sidelobes across the entire sky, meaning almost all sources are blended with one another. Deconvolving this PSF correctly is a challenging task. The \clean algorithm is one approach to performing the deconvolution. However, as discussed in \cref{Introduction}, \clean is highly interactive and non-linear, meaning that by the time a \clean image has been produced, the effective PSF in the image is some unknown function, usually modelled as a Gaussian fit to the central lobe of the dirty beam (and known as the `\clean beam'). 

For source morphology this leads to un-modelled shape distortions due to the PSF which are not captured by the \clean beam. To account for this additional distortion, we use simulations of the VLA COSMOS dataset to estimate an effective PSF model in the image plane. We additionally use these simulations to probe for shape systematic effects induced by the imaging pipeline.

The input sky image used for the simulation was created with the GalSim (modular galaxy image simulation) toolkit \citep{2015A&C....10..121R}. Our simulations include the positions and morphologies of all sources identified in the source catalogue constructed by \cite{2007ApJS..172...46S}. We note that it is primarily the distribution of source positions, fluxes and sizes which determines the characteristics of the un-\clean-ed flux contributing to the residual distortions in the effective PSF in the image. By using the \cite{2007ApJS..172...46S} catalogue as input to our simulations, we ensure that the position, flux and size distributions of sources in our simulations are approximately the same as the true distributions in the COSMOS field. The catalogue of \cite{2007ApJS..172...46S} was constructed using the \textsc{aips} tasks `\textsc{sad}'\footnote{`\textsc{Sad}' (Search and Destroy) is a source extraction program within \textsc{aips}.}, `\textsc{maxfit}'\footnote{`\textsc{Maxfit}' is a coordinates and value extremum finder algorithm within \textsc{aips}, useful for finding the centre position of sources.} and `\textsc{jmfit}'\footnote{`\textsc{Jmfit}' is a Gaussian fitting algorithm within \textsc{aips}.} to determine the source position, flux density and morphologies. 

We generate the resolved sources in our simulations assuming elliptical Gaussian morphologies, and using the best-fit parameters as determined by \cite{2007ApJS..172...46S}. For all sources that were classified as unresolved by \cite{2007ApJS..172...46S}, we inject a single pixel source (delta function) with the appropriate flux and position into the simulation. The simulations are created on a real-space grid with a pixel scale of 0.2$^{\prime\prime}$/pixel, which is higher resolution than that of the final radio image which is gridded to 0.35$^{\prime\prime}$/pixel. For each of the 23 COSMOS field pointings we project this simulation onto the visibilities corresponding to the true Fourier plane sampling of the original VLA dataset. Finally, random noise (with RMS equal to that measured from the real visibility data) is added to the simulated visibilities. 

The simulated visibility data sets thus contain the same dirty beam, the same level of random noise, and approximately the same source distribution as was found in the original data set. Additionally, the resolved source morphologies are exactly known and can serve as a test of the imaging pipeline. The simulated datasets were then processed using the exact same imaging pipeline as described in \cref{Imaging}, using the \textsc{casa} task \clean. The output of our simulation pipeline is thus a re-constructed \clean image which should fully capture any systematic effects introduced by the radio imaging process.

By examining the observed shapes of those objects that were injected as point sources, we were then able to estimate the effective PSF across the image. In order to capture any spatial variation in the PSF, the field was split into a $4\times4$ grid and a weighted mean of the measured shapes corresponding to the injected point sources was calculated in each of the $4\times4$ grid cells. We used the signal to noise ratio ($S/N$) as the weight for each source, 
\begin{equation}
\text{PSF}_{\text{Total}} = \frac{1}{\sum\limits_{i} \omega_{i}}  \sum_{i} \omega_{i} I_{i},
\label{eq:PSF}
\end{equation}
where $\omega_{i} = (S/N)_{i}$. We found that the resulting weighted mean beam shapes in each grid cell were well modelled using 2D Gaussian ellipses though the PSF shape did vary somewhat across the field. 
Note that we extracted estimates of the PSF from the simulations for the natural and uniform weighting schemes separately. Cut-outs of the final PSF model images were produced with a $32\times32$ stamp size (equivalent to a $11.2\times11.2$ arcseonds), and were later up-sampled\footnote{The original cut-outs were upsampled from the original sidelength of $s = 32$ pixels to an upsampled sidelength $S = 170$ pixels, using a padding co-efficient of $p=2$ and an up-sampling coefficient of $u=3$, following the relation $S=(s+p) \times u$.} to $170\times170$ cut-outs. These PSF cut-outs were subsequently used in the shape analysis of both the resolved simulated sources, and of the resolved real sources.  

\begin{figure*}
 \centering
 \includegraphics[width=0.55\textwidth]{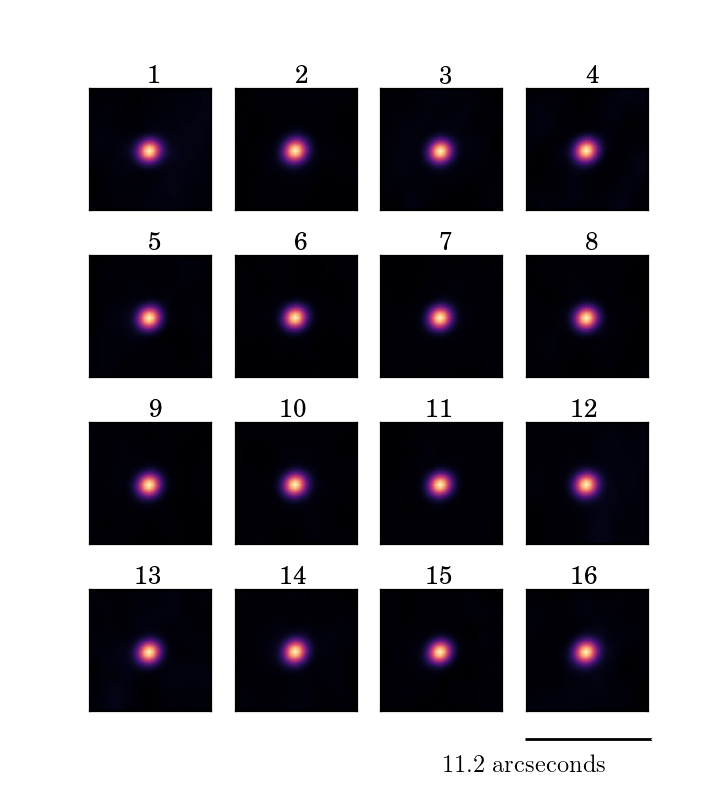}

 \caption[Fitted Sectional PSF Models]{The spatially-dependent PSF of the VLA-COSMOS image as estimated from our simulated datasets. The full VLA-COSMOS field was split into 16 equal-size regions and the images show the average PSF within each section. This figure shows the PSF corresponding to the naturally-weighted image -- the PSF corresponding to the uniformly-weighted image is broadly similar. These mean recovered PSFs are subsequently fitted with an elliptical Gaussian model to provide an effective PSF model for each region on the radio mosaic. The grid numbers shown in bold black above each cutout correspond to the grid numbers reported in \cref{Tab:SectionalPSF}. Each plot shares a linear colour scale}\label{fig:FittedPSFsSectional}
\end{figure*}

\cref{fig:FittedPSFsSectional} displays the weighted mean beam shapes as measured from our simulations for the case of natural weighting (a similar equivalent can be made for the uniform weighting scheme). The best-fit parameters of the 2D elliptical Gaussian fits are provided in \cref{Tab:SectionalPSF} for both weighting schemes. Since the effective beam size measured shows some dependency on grid position, this will be an important feature to include in our shape analysis of the real data.
\begin{table*}
\centering
\begin{tabular}{l|cccc|cccc}
\hline
   & \multicolumn{4}{|c|}{Natural Weighting} & \multicolumn{4}{|c}{Uniform Weighting}\\\cmidrule{2-9}
  Grid Number &  $a$ [arcsec] & $b$ [arcsec] & $\alpha$ [deg] & $\boldsymbol{e}$ & $a$ [arcsec] & $b$ [arcsec] & $\alpha$ [deg] & $\boldsymbol{e}$ \\
 \hline
  1 & 1.404 & 1.385 & 22.1 & 0.0135 & 1.320 & 1.235 & 19.4 & 0.0664\\
  2 & 1.449 & 1.449 & 45.0 & 0.0001 & 1.386 & 1.314 & 25.6 & 0.0536\\
  3 & 1.386 & 1.385 & 41.8 & 0.0009 & 1.357 & 1.301 & 16.5 & 0.0419\\
  4 & 1.411 & 1.409 & 39.5 & 0.0012 & 1.330 & 1.287 & 23.5 & 0.0335\\
  5 & 1.380 & 1.371 & 37.4 & 0.0066 & 1.336 & 1.234 & 20.8 & 0.0780\\
  6 & 1.381 & 1.339 & 33.2 & 0.0311 & 1.327 & 1.250 & 18.5 & 0.0589\\
  7 & 1.388 & 1.383 & 37.6 & 0.0037 & 1.327 & 1.253 & 17.6 & 0.0574\\
  8 & 1.396 & 1.361 & 33.2 & 0.0257 & 1.315 & 1.250 & 18.5 & 0.0509\\
  9 & 1.399 & 1.377 & 38.8 & 0.0157 & 1.314 & 1.240 & 22.4 & 0.0583\\
  10 & 1.410 & 1.396 & 33.6 & 0.0096 & 1.353 & 1.286 & 17.4 & 0.0504\\
  11 & 1.372 & 1.363 & 39.5 & 0.0068 & 1.324 & 1.243 & 21.4 & 0.0635\\
  12 & 1.412 & 1.386 & 34.3 & 0.0190 & 1.347 & 1.265 & 17.7 & 0.0628\\
  13 & 1.363 & 1.329 & 36.6 & 0.0247 & 1.358 & 1.278 & 21.5 & 0.0604\\
  14 & 1.428 & 1.427 & 47.3 & 0.0009 & 1.350 & 1.298 & 32.4 & 0.0394\\
  15 & 1.376 & 1.373 & 39.9 & 0.0021 & 1.323 & 1.255 & 25.0 & 0.0530\\
  16 & 1.454 & 1.450 & 39.7 & 0.0031 & 1.377 & 1.317 & 21.9 & 0.0443\\
 \hline
 Mean & 1.401 & 1.386 & 37.4 & 0.0103 & 1.340 & 1.269 & 21.2 & 0.0546 \\
 $\sigma$ & 0.026 & 0.033 & 5.5 & 0.0098 & 0.021 & 0.027 & 3.9 & 0.0109\\
 \hline
\end{tabular}
\caption[Fitted Sectional PSF Model Parameters]{The parameter outputs from the 2D Gaussian model fits are provided for both the naturally and uniformly weighted radio images. We quote the grid position, the semi-major ($a$) and semi-minor ($b$) axes, the position angle ($\alpha$, measured in degrees North of East) and the ellipticity modulus ($\boldsymbol{e}$). The bottom two rows report the mean and standard deviation ($\sigma$) for each column.} \label{Tab:SectionalPSF}
\end{table*}

\subsection{Estimation of radio shapes}
\label{radio_shape_pipeline}
Having estimated the effective PSF across the image, we are now in a position to measure the shapes of those sources that are resolved by the radio observations. Here we outline our general approach to extracting source shapes. We apply this approach to the resolved sources in our simulations, and to the resolved sources in the real data, in Sections \ref{simulation_err} and \ref{vla_shape_measurements} respectively. When fitting for shapes, we take the source positions from \cite{2007ApJS..172...46S} who performed their source finding with the \textsc{aips} task \textsc{sad}. We then classify each source as resolved/unresolved according to an elliptical Gaussian fitting procedure. We use the \textsc{casa} task \textsc{imfit} (which incorporates the deconvolution of the \clean beam) to fit elliptical Gaussian models to each source. We classify as resolved, those sources with a major and minor axis greater than $\sim1.2$ times the effective PSF size. For the natural weighted image, we select sources according to:
\begin{align*}
a &> 1.8\ \mathrm{arcseconds} \\
b &> 1.7\ \mathrm{arcseconds} \\
S/N &> 5,
\end{align*}
whereas for the uniform-weighted image:
\begin{align*}
a &> 1.75\ \mathrm{arcseconds} \\
b &> 1.6\ \mathrm{arcseconds} \\
S/N &> 5,
\end{align*}
where $a$ and $b$ are the best-fit semi-major and semi-minor axes respectfully. A $S/N$ cut of $> 5$ was applied, as was previously in \cite{2007ApJS..172...46S}. Since we are only concerned in position angles for this study we deem this relatively low $S/N$ cut as sufficient to maximise source counts, although it should be noted that typical weak lensing studies would require a much larger cut and therefore reducing the sample size. After applying these cuts, we are left with $931$ and $762$ remaining sources in the natural and uniform weighted images respectfully.

\subsubsection{Shape Estimator}
In this work we make use of developments in optical shape measurement techniques. In particular, we have chosen to use the \textsc{im3shape} code to extract galaxy shapes, which we apply directly to our radio image. \textsc{Im3shape} \citep{2013MNRAS.434.1604Z} is a maximum likelihood model-fitting algorithm for estimating galaxy ellipticities. The performance of \textsc{im3shape} was demonstrated, for example, during the GREAT10 optical weak lensing challenge \citep{2012MNRAS.423.3163K}, where it achieved relatively accurate and unbiased shape measurements \citep{2012MNRAS.427.2711K}.

We parametrise individual galaxy shapes in terms of a two-component ellipticity, $\boldsymbol{e} = (e_1, e_2)$. Here $e_1$ describes elongations parallel and perpendicular to an arbitrarily chosen reference axis and $e_2$ describes elongations along the directions rotated $\pm 45^{\circ}$ from the reference axis. This ellipticity can also be expressed in terms of a semi-major ($a$) and semi-minor ($b$) axis, and a position angle ($\alpha$). The ellipticity modulus $e = |\boldsymbol{e}|$ is given by
\begin{equation}
e = \frac{a^{2}-b^{2}}{a^{2}+b^2} = \sqrt{\mathit{e}^{2}_{1} + \mathit{e}^{2}_{2}},
\label{eq:1.4}
\end{equation}
where $\mathit{e}_{1}$ and $\mathit{e}_{2}$ are the vector elements of $\mathit{e}$,
\begin{equation}
\mathit{e}_{i} = \begin{pmatrix}  |\boldsymbol{e}|  \hspace{1. mm}\cos(2 \alpha)  \\
   |\boldsymbol{e}|  \hspace{1. mm}\sin(2 \alpha)\end{pmatrix} ,
\label{eq:1.5}
\end{equation}
and where $\alpha$ is the position angle, 
\begin{equation}
\alpha = \frac{1}{2} \tan^{-1}\left(\frac{\mathit{e}_{2}}{\mathit{e}_{1}} \right).
\label{eq:pa}
\end{equation}

In our shape measurement analysis, we used \textsc{Im3shape} to fit a sum of two S{\`e}rsic profiles, specifically a ``Bulge plus Disc" model. We emphasise that the \textsc{Im3shape} fitting procedure fully accounts for the PSF (which the user must supply) by including it in the forward model. A S{\`e}rsic profile is a surface brightness distribution model described by,
\begin{equation}
I(r) = I(0)\exp\left[-\left(\frac{r}{r_{0}}\right)^{\frac{1}{n}}\right],
\label{eq:sersic}
\end{equation}
where $I(0)$ is the central intensity, $r$ is the projected radius, $r_{0}$ is the and scale length of the profile and $n$ is the S{\`e}rsic index \citep{2001MNRAS.321..269T}. The maximum likelihood approach optimises the model parameters by minimising the summation of the square differences between the model and the data with each iteration, i.e. it minimises the $\chi^{2}$ misfit statistic. 

We provide \textsc{im3shape} with galaxy positions (from the \citealt{2007ApJS..172...46S} source catalogue) and PSF models at each of those positions (approximated by the 16 PSF models derived from simulations, see section \ref{Simulation}). \textsc{Im3shape} preforms a fit for source parameters describing the position and shape of the galaxy in terms of the two-component S{\`e}rsic Bulge + Disc model. Although it increases the computational time from the default \textsc{im3shape} settings, we choose to keep both the Disc and Bulge components of the galaxy models' S{\`e}rsic index free. This is motivated in part since our sample is sufficiently small that this addition of computation time is not problematic. Moreover the standard S{\`e}rsic profile, which is well tested in comparison to real optical galaxies lacks the same test in comparison to galaxies as observed in the radio. By leaving the two indices free we allow for a larger range of models. This shape measurement procedure is performed for the entire galaxy population, and for the two weighting schemes that we have implemented.  

\section{Quality of Shape measurements from Simulations}\label{simulation_err}

In this section, we use the simulations described in \cref{Simulation} to quantify how well the galaxy shape measurement pipeline of \cref{radio_shape_pipeline} can recover the input ellipticities.

\subsection{Estimation of measurement bias}
\label{sim_bias_estimation}

\subsubsection{Ellipticity measurement bias}


To process our simulations, we mirror the analysis performed on the real VLA COSMOS data, producing a recovered image from the simulated visibilities using the \clean algorithm as described in \cref{Imaging}, and measuring the shapes of the galaxies from the recovered image using \textsc{IM3SHAPE}.

We quantify the success of the shape recovery from the simulations in terms of a linear bias model. This is parametrised for each ellipticity component separately as,
\begin{equation}
\mathit{e}^{\mathrm{obs}} = (1+m)\mathit{e}^{\mathrm{true}} + c ,
\label{eq:linear_bias_model}
\end{equation}
where $m$ is a multiplicative, and $c$ an additive, bias.

\begin{figure*}
    \centering
    \begin{subfigure}[b]{0.475\textwidth}
           \centering
           \includegraphics[width=\textwidth]{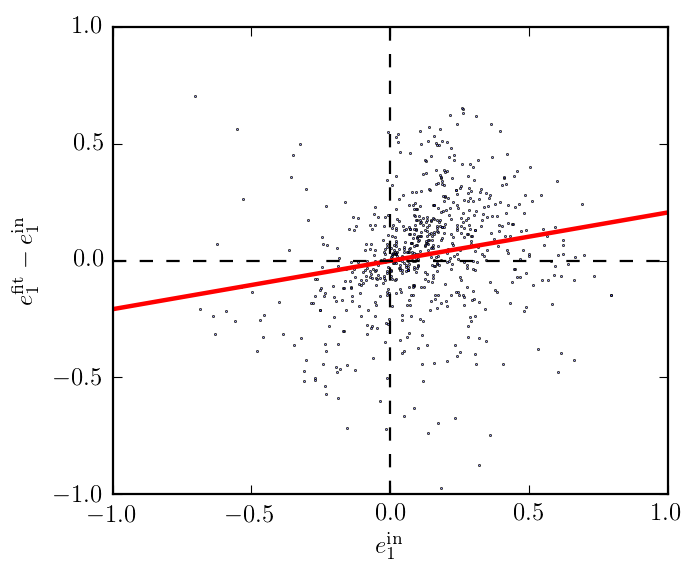}
            \caption{Natural weighting, $\mathit{e}_{1}$ parameter recovery.}
            \label{fig:radioe1recoverya}
    \end{subfigure}
    \begin{subfigure}[b]{0.475\textwidth}
            \centering
            \includegraphics[width=\textwidth]{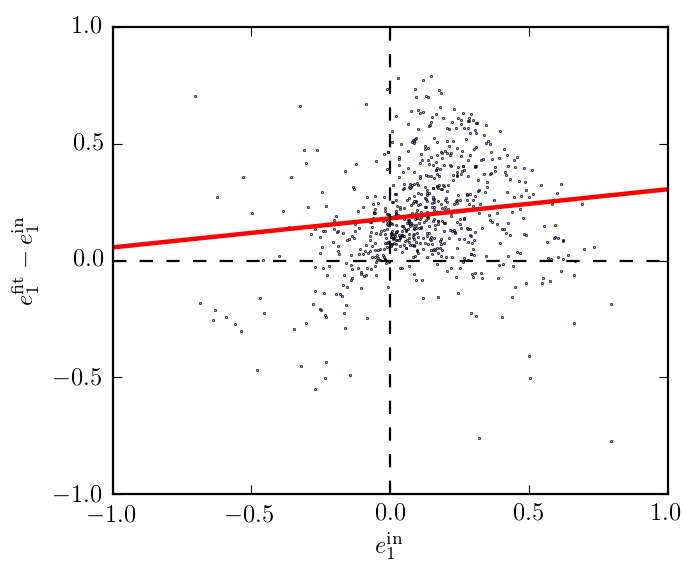}
            \caption{Uniform weighting, $\mathit{e}_{1}$ parameter recovery.}
            \label{fig:radioe1recoveryb}
    \end{subfigure}
    \vskip\baselineskip
    \begin{subfigure}[b]{0.475\textwidth}
           \centering
           \includegraphics[width=\textwidth]{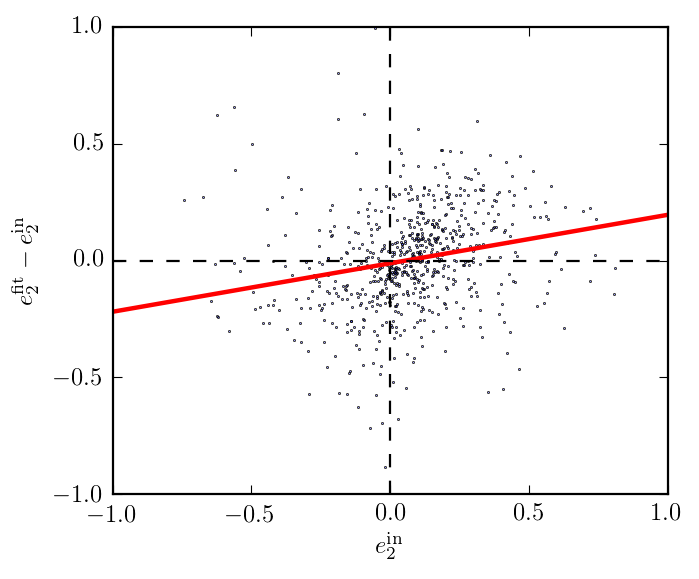}
            \caption{Natural weighting, $\mathit{e}_{2}$ parameter recovery.}
            \label{fig:radioe2recoverya}
    \end{subfigure}
    \begin{subfigure}[b]{0.475\textwidth}
            \centering
            \includegraphics[width=\textwidth]{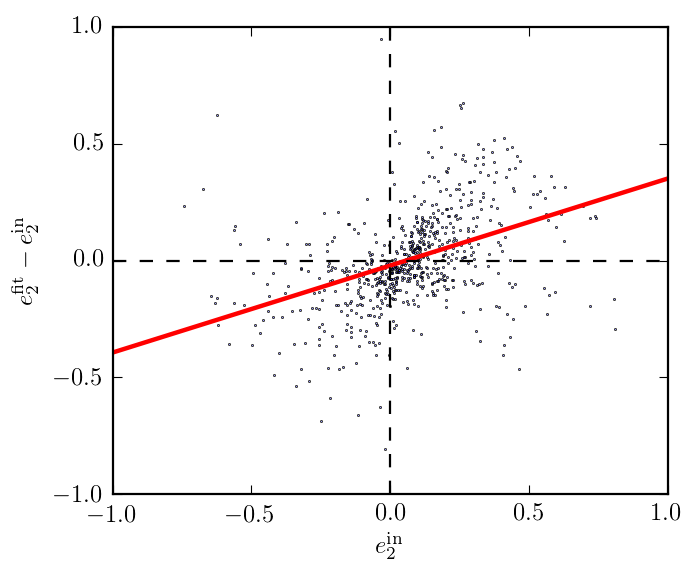}
            \caption{Uniform weighting, $\mathit{e}_{2}$ parameter recovery.}
            \label{fig:radioe2recoveryb}
    \end{subfigure}
    \caption[Parameter Bias]{Shape parameter recovery from simulations. The figures show the residuals in the ellipticity components as measured from the simulated radio datasets. Each point represents the measurements for a single galaxy. Panels (a) \& (c) show the recovery of $\mathit{e}_{1}$ and $\mathit{e}_{2}$ from images produced using natural weighting. Panels (b) \& (d) show the same for images produced using uniform weighting. In each panel, a regression line is overplotted using the best fitting parameters reported in \cref{Tab:m_and_c}.}\label{fig:radioerecovery}
\end{figure*}

\cref{fig:radioerecovery} shows the recovery of the input $\mathit{e}_{1}$ and $\mathit{e}_{2}$ values for both weighting choices in the imaging algorithm, with the best-fitting multiplicative and additive bias parameters listed in \cref{Tab:m_and_c}. For the natural weighting scheme we find a relatively small additive bias and the multiplicative bias appears highly similar between the two ellipticity components. However, this is not true for the uniform weighting scheme, which displays a more significant additive systematic effect and a clear discrepancy in the multiplicative bias components.

\begin{table}
\centering
\resizebox{0.475\textwidth}{!}{
\begin{tabular}{lccc}
\hline
 & Parameter & $m$ & $c$ \\
\hline
 & $\mathit{e}_{1}$ & $0.207\pm 0.042$ & $-0.001\pm 0.011$ \\
 Natural & $\mathit{e}_{2}$ & $0.208\pm 0.035$ & $-0.012\pm 0.008$ \\
 Weighting & $\alpha^{\mathrm{raw}}$ & $-0.0022\pm 0.0116$ & $-0.0099\pm 0.0080$\\
 & $\alpha^{\mathrm{corr}}$ & $0.0115\pm 0.0116$ & $-0.0118\pm 0.0079$ \\
\hline
 & $\mathit{e}_{1}$ &  $0.125\pm0.039$ & $0.181\pm0.010$\\
 Uniform & $\mathit{e}_{2}$ &  $0.373\pm0.032$ & $-0.021\pm0.008$\\
 Weighting & $\alpha^{\mathrm{raw}}$ & $-0.0870\pm 0.0081$ & $-0.0058\pm 0.0056$\\
 & $\alpha^{\mathrm{corr}}$ & $0.0247\pm 0.0096$ & $0.0120\pm 0.0066$ \\
\hline
\end{tabular}
}
\caption[Fitted Bias from Simulation]{Best-fitting bias parameters for ellipticity components ($\mathit{e}_{1}$ and $\mathit{e}_{2}$), and for both the raw and corrected position angle components ($\alpha^{\mathrm{raw}}$ and $\alpha^{\mathrm{corr}}$; see text in \cref{sim_bias_estimation}) as measured from simulations. The measured ellipticity and position angle components are fit to a linear bias model with multiplicative ($m$) and additive ($c$) bias parameters. $\alpha^{\mathrm{raw}}$ refers to position angles estimated from the measured ellipticities using \cref{eq:posn_angle_def}. $\alpha^{\mathrm{corr}}$ refers to position angles estimated from ellipticities which have already been corrected for the measured $\mathit{e}_{i}$ biases.}\label{Tab:m_and_c}
\end{table}

\begin{figure*}
    \centering
    \begin{subfigure}[b]{0.475\textwidth}
           \centering
           \includegraphics[width=\textwidth]{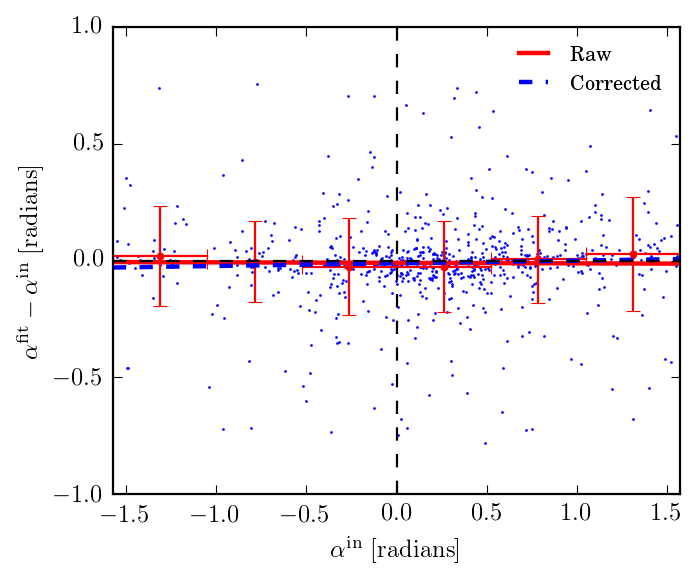}
            \caption{Natural weighting, $\alpha$ parameter recovery.}
            \label{fig:radioalpharecoverya}
    \end{subfigure}
    \begin{subfigure}[b]{0.475\textwidth}
            \centering
            \includegraphics[width=\textwidth]{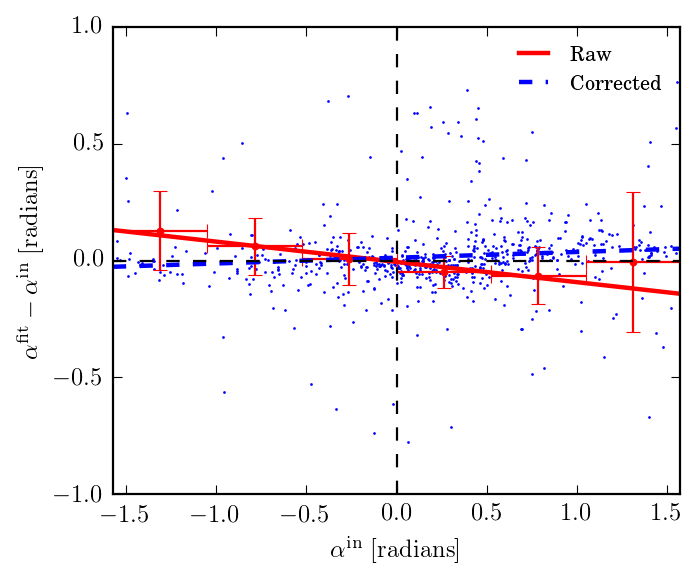}
            \caption{Uniform weighting, $\alpha$ parameter recovery.}
            \label{fig:radioalpharecoveryb}
    \end{subfigure}
    \caption[Parameter Bias]{Residuals in the recovery of the galaxy position angles ($\alpha$) for both weighting schemes considered (\emph{left}: natural weighting, \emph{right}: uniform weighting). The points show the bias corrected points while the binned error bars show the distribution before bias correction. In addition to the ``raw" position angle estimates, the figure also shows the result for the bias-corrected measurements as described in \cref{shear_bias_from_alpha_bias}. For each analysis, a regression line is over-plot: red (solid) and blue (dashed) lines show the regression line fits before and after bias corrections respectfully.}\label{fig:radioalpharecovery}
\end{figure*}

\subsubsection{Position angle measurement bias}\label{Position angle measurement bias}

In addition to measuring the accuracy of the ellipticity reconstruction, we have also assessed the accuracy with which the galaxy position angles can be reconstructed. The position angle of a galaxy can be estimated from measurements of its ellipticity components as,
\begin{equation}
\alpha^{\mathrm{obs}} = \frac{1}{2} \tan^{-1} \left(\frac{\mathit{e}_{2}^{\mathrm{obs}}}{\mathit{e}_{1}^{\mathrm{obs}}}\right).
\label{eq:posn_angle_def}
\end{equation}
From \cref{eq:linear_bias_model} it is clear that, for a small additive ellipticity bias and highly similar multiplicative ellipticity biases, position angle estimates based on \cref{eq:posn_angle_def} should be relatively unbiased. We therefore expect to recover relatively unbiased position angle estimates from our naturally weighted images for which we found small additive ellipticity biases and highly similar multiplicative biases.  

\cref{fig:radioalpharecovery} shows the position angles $\alpha^{\mathrm{obs}}$, recovered from the source ellipticities measured in our simulations. In addition to the ``raw" position angles $\alpha^{\rm raw}$ we have also estimated the position angles from ellipticity measurements which have already been debiased using the best-fitting linear ellipticity bias model parameters. We label these corrected position angle measurements $\alpha^{\rm corr}$. We fit a linear bias model to both the $\alpha^{\rm raw}$ and $\alpha^{\rm corr}$ estimates (as a function of the true position angles $\alpha^{\mathrm{true}}$). \cref{fig:radioalpharecovery} shows that the extracted position angles using the natural weighting scheme are essentially unbiased, both before and after correction for the measured ellipticity bias, although they are still a noisy estimate. The same is not true for the uniform weighting scheme where a residual multiplicative bias remains. Based on these results, we choose the natural weighting scheme for the remainder of this work. Note that although we have used a simple linear model for the bias in $\alpha$, this may not necessarily be a good model in the limit of very precise measurements. However, it appears to provide a good approximation given the level of measurement noise present here. The best-fitting parameters for this linear model are listed in \cref{Tab:m_and_c}. The associated uncertainties from values in \cref{Tab:m_and_c} were obtained from the covariance matrix of the linear model fit.

\begin{figure*}
    \centering
    \begin{subfigure}[b]{0.475\textwidth}
           \centering
           \includegraphics[width=\textwidth]{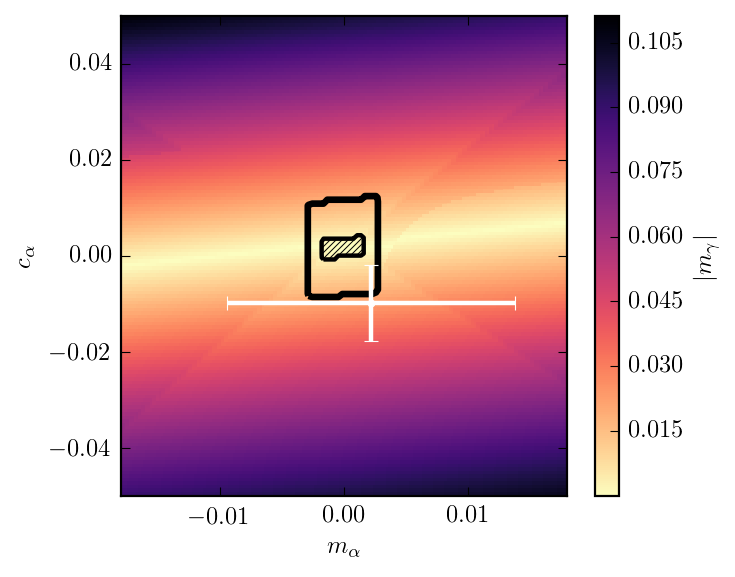}
            \caption{Absolute $m_{\gamma}$ colour scale.}
            \label{fig:bias_linear_m__finer_colour_contour}
    \end{subfigure}
    \begin{subfigure}[b]{0.475\textwidth}
            \centering
            \includegraphics[width=\textwidth]{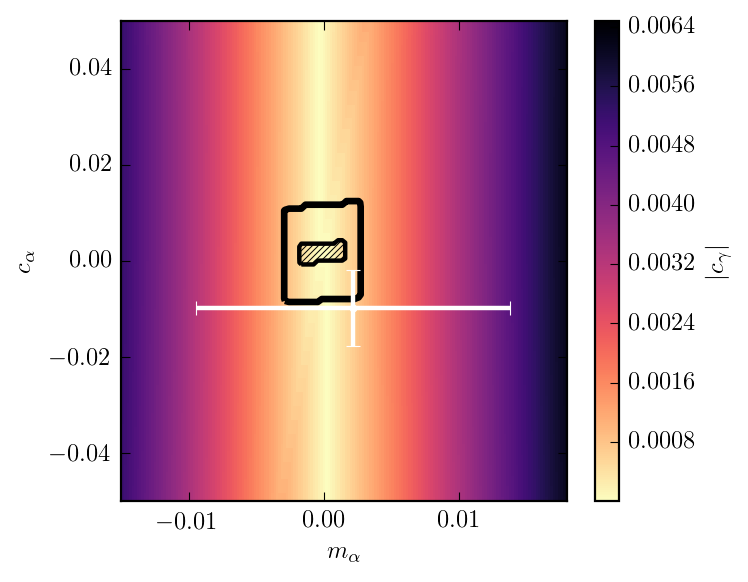}
            \caption{Absolute $c_{\gamma}$ colour scale.}
            \label{fig:bias_linear_c__finer_colour_contour}
    \end{subfigure}
    \caption[$\alpha$ Target Contours]{Relation between position angle bias and lensing shear bias as derived from simulations. The colour scale shows the level of multiplicative shear bias ($|m_\gamma|$, \emph{left panel}) and additive shear bias ($|c_\gamma|$, \emph{right panel}) as a function of the position angle bias ($m_\alpha$ and $c_\alpha$) injected into the simulations. For comparison, the requirements on $m_\gamma$ and $c_\gamma$ for Stage II and Stage III surveys from \cref{Tab:m_and_c_shear} are shown as the area enclosed by the outer and inner contours respectively. The performance of our position angle estimator, as derived from the VLA-COSMOS simulations of \cref{sim_bias_estimation}, is over-plotted in each panel as a single point with error bars. For further detail see \cref{shear_bias_from_alpha_bias}.}  
    \label{fig:radioarecovery}
\end{figure*}

In \cref{shear_bias_from_alpha_bias} we investigate how the a position angle recovery obtained may relate to requirements on shear additive and multiplicative systematics in current and future optical weak lensing surveys. \cref{fig:radioarecovery} demonstrates the shear recovery achievable using a position angle-only shear estimators (as detailed in \cite{2014MNRAS.445.1836W}). For comparison we have over-plotted with the white error bar the approximate results obtained in from simulations in this study, additionally we show weak lensing requirements with optical studies with the black contours. We can see here that although our position angle estimates are relatively unbiased within measurement errors, the resultant uncertainty on the position angle bias parameters from this study is large compared to current capabilities. For more information on \cref{fig:radioarecovery} see \cref{shear_bias_from_alpha_bias}.

\section{Radio Shape Measurements with Data}\label{vla_shape_measurements}
For the radio images we have constructed from the real data (with the pipeline described in \cref{Imaging}), we measure source ellipticities with \textsc{im3shape}, with the application of simulated effective PSF models approximated over the field (see \cref{Simulation}). We convert these ellipticity measurements to position angle estimates, $\alpha$ using \cref{eq:pa} to create the full galaxy morphology catalogue\footnote{A source catalogue can be found at \url{https://github.com/bentunbridge/cosmos-radio-optical.}}.
Note that the $\mathit{e}_{i}$ components are corrected for multiplicative and additive bias, as estimated from the corresponding simulations, following \cref{eq:linear_bias_model}. However, this has little effect on the resulting $\alpha$ values for the image created with natural weighting (see discussion in \cref{simulation_err}).

\begin{figure}
 \centering
 \includegraphics[width=0.475\textwidth]{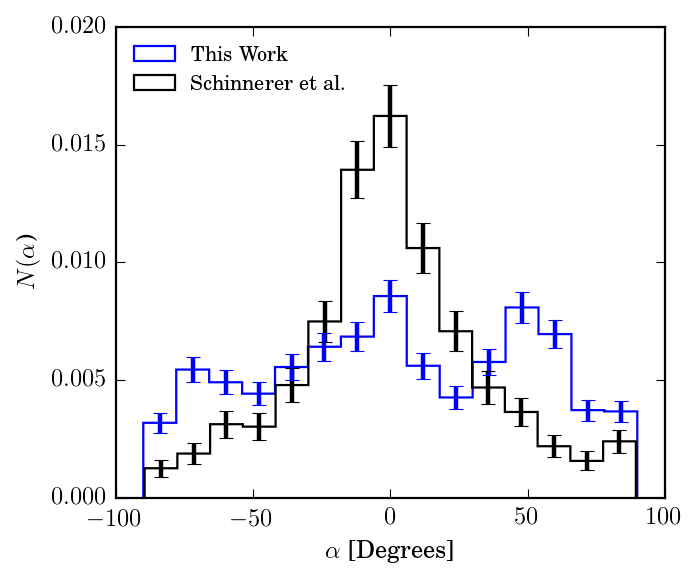}
 \caption[Radio $\alpha$ parameter compared]{Normalised distribution of source position angles, $\alpha$, as estimated from two independent imaging analyses of the VLA COSMOS data. The black points show the distribution based on the image produced by \cite{2007ApJS..172...46S}, while the blue points show the distribution from the image created in this work using natural weighting. Poisson error bars are included.}
 \label{fig:radioPA_compare}
\end{figure}

As a measure of the improvement in our VLA COSMOS imaging (from the point of view of shape estimation) over the original analysis of \cite{2007ApJS..172...46S}, in \cref{fig:radioPA_compare} we compare histograms of the measured position angles from the two analyses. The figure reveals a large discrepancy in the distributions measured from the two images, with the position angles extracted from our new image being distributed much closer to the uniform distribution expected for such a large sample. The large systematic correlations in position angles present in the original image would clearly pose a problem for any subsequent analyses which rely on accurate measurements of galaxy shapes. 

We compare the distribution of the galaxy ellipticity moduli, $|\boldsymbol{e}|$, measured from our new VLA COSMOS image, to a model distribution in \cref{fig:E_hist_opt}. The model we compare to is a reasonable approximation of a typical optical shape distribution, and is taken from the GRavitational lEnsing Accuracy Testing 2008 (GREAT08) analysis \citep{2010MNRAS.405.2044B}. The functional form is given by
\begin{equation}\label{eq:STEP}
    P(|\boldsymbol{e}|) = |\boldsymbol{e}| \left[\cos\left(\frac{\pi |\boldsymbol{e}|}{2}\right)\right]^{2} \exp\left[-\left(\frac{2|\boldsymbol{e}|}{B}\right)^{C}\right],
\end{equation}
where parameters, $B = 0.19$ and $C = 0.58$ are appropriate for disc-dominated galaxies. We find that the $|\boldsymbol{e}|$ distribution measured from the radio data is slightly broader than that expected for an optical sample. This may be a genuine feature of the shape population or a possible effect in the radio imaging that has not been accounted for. We measure an ellipticity dispersion of $\sigma_{e}$ = $0.29$ (per ellipticity component) and best-fitting model parameter values of B = $0.113\pm 0.041$ and C = $0.303\pm 0.058$ in \cref{eq:STEP}. In \cref{fig:E_hist_opt}, we also show the distribution recovered from our analysis of the HST optical data (see next Section), finding a much better agreement with the model curve assuming the fiducial parameters (as expected). For the optical data we find best-fitting model parameter values of $B = 4.65\pm{0.288\E^{-5}}$ and $C = 0.218\pm0.002$.


\section{Multi-Waveband Shape Comparison}\label{Multi-Waveband Shape Comparison}
Weak lensing at radio wavelengths can provide an important complementary study to equivalent studies at optical wavelengths. An important aspect of such complementarity is how radio and optical galaxy shapes intrinsically relate to each other. This has important implications for how a radio weak lensing survey may be used in synergy with overlapping optical observations. For example if radio and optical shapes are typically intrinsically aligned then the radio provides an alternative measure, with different systematics, but probing the same weak lensing sample as the optical observations. On the other hand if radio and optical shapes are intrinsically uncorrelated then the radio effectively probes a different sample of galaxies, at the same redshifts as the optical catalogue, but again with different systematics. In \cref{fig:radioalpharecovery} we demonstrated that our position angle estimator, though noisy, provided unbiased measurements of the galaxy position angles from the simulated radio datasets. We therefore conduct our galaxy shape comparative study in the COSMOS field using galaxy position angle measurements.

\subsection{Optical shape measurement}
We measured the shapes of galaxies in the optical from the HST-ACS data described previously in \cref{HST data}. The high density of sources in the HST image allows for PSF models to be created from the sample of stars detected across each CCD tile exposure. Stars, unlike galaxies, should be intrinsically point-like, and so can be used to estimate the PSF over the observed field. For this task we use the \textsc{PSFex} (`PSF Extractor') software \citep{2011ASPC..442..435B} which has been used in previous optical weak lensing studies \citep[e.g.][]{2015arXiv150705603J}.

\begin{figure}
 \centering
 \includegraphics[width=0.5\textwidth,angle=0]{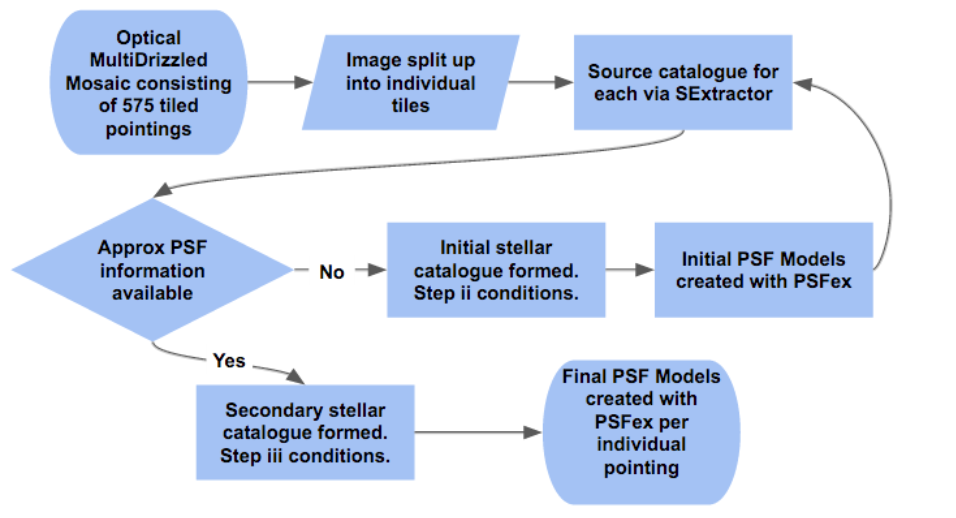}
 \caption[Optical PSF Modelling Pipeline]{A flow chart depicting the optical PSF modelling pipeline. Each of the 575 pointings are first run through \textsc{SExtractor} and \textsc{PSFex} to form PSF models with an initial stellar population estimate. Following this an improved stellar population is extracted and subsequently improved \textsc{PSFex} models are generated.}
 \label{fig:Optical PSF Pipeline}
\end{figure}

To apply \textsc{PSFex} to the HST data, we follow the pipeline used for the weak lensing analysis of the DES data \citep{2015arXiv150705603J}, which we summarize here.  
\begin{enumerate}
\item \label{n1}We create a source catalogue with the source extraction software, \textsc{SExtractor} \citep{1996A&AS..117..393B}. We do this individually for each observation pointing with the aid of a corresponding RMS noise map for the observation. For these observations, the F814W magnitude zero-point is set to $27.2$ to match the limiting point-source depth achieved in the observations \citep{2007ApJS..172..196K}.


\item \label{n2}A catalogue of stars is required for each field in order to form \textsc{PSFex} models. For the first iteration of the models, we apply a selection cut of \textsc{class\_star} $> 0.6$ to the catalogues produced in \ref{n1}, where \textsc{class\_star} is a star/galaxy classification parameter and described in more detail in \cite{1996A&AS..117..393B}. Additionally only stars with $S/N$ $>$ 20 are considered in this sample. This provides us with a sample that is mostly composed of stars and point-like objects, and serves as a rough basis to form \textsc{PSFex} models. The reliability of \textsc{class\_star} is discussed in \cite{2005astro.ph.12139H}.

\item \label{n3} We conduct a second iteration of star classification on the data. Once again we use \textsc{SExtractor} to produce a source catalogue, where the software is now provided with the PSF models produced in step \ref{n2} with \textsc{PSFex}. For this iteration, the software incorporates the PSF model into the extracted parameters and we are able to produce a more reliable star/galaxy classification. We adapt the pseudo-code detailed in \cite{2015arXiv150705603J} for our own star source selection which we implement as follows:\\
\newline
\begin{minipage}{0.7\linewidth}
\begin{small}
\begin{tabular}{p{1.65cm}p{0.2cm}p{5cm}}
\textsc{bright\_test} &$=$& \textsc{class\_star} $> 0.6$\\
 &&\textsc{and} \textsc{mag\_auto} $< 25.0$\\
 &&\\
\textsc{locus\_test} &$=$& \textsc{spread\_model}\\
 &&$+$\textsc{spreaderr\_model} $< 0.003$\\
 &&\\
\textsc{faint\_psf\_test} &$=$& \textsc{mag\_psf} $> 40.0$\\
 &&\textsc{and} \textsc{mag\_auto} $< 26.0$\\
 &&\textsc{and} \textsc{mag\_psf} $< 90.0$\\
 &&\\ 
\textsc{galaxies} &$=$& \textsc{not} \textsc{bright\_test}\\
 &&\textsc{and} \textsc{not} \textsc{locus\_test}\\
 &&\textsc{and} \textsc{not} \textsc{faint\_psf\_test}\\
 &&\\ 
\textsc{stars} &$=$& [\textsc{locus\_test} \textsc{or} \textsc{bright\_test}]\\
 &&\textsc{and} \textsc{not} \textsc{faint\_psf\_test}\\
\end{tabular}
\end{small}
\end{minipage}\\
\newline

Sources are only included in our star sample if they are identified by the \textsc{bright\_test} or \textsc{locus\_test} and if they are not removed as junk in the \textsc{faint\_psf\_test}. Here the \textsc{bright\_test} procedure simply relies on \textsc{SExtractor}`s own source classification system, this time excluding bright sources (Magnitude $>25$) from the sample. The \textsc{locus\_test} procedure identifies the sourc locations relative to the stellar locus via the \textsc{spread\_model} parameter. This is an additional star/galaxy classifier based on the linear discriminant between the best fitting PSF model (from step \ref{n2}) and a model made from the same PSF convolved with a circular exponential disc model \citep[see][for a more detailed discussion on the \textsc{spread\_model} classifier]{2012ApJ...757...83D}. Again we only include stars with $S/N>20$ in our final stellar sample. 
\item \label{n4} Finally we again generate PSF models, this time from the improved stellar sample from step \ref{n3} and using \textsc{PSFex}. This provides a PSF model as a function of position on the sky for each of the 575 pointings across the entire mosaic, which we will use for shape analysis.
\end{enumerate}
These steps are depicted as a flowchart in \cref{fig:Optical PSF Pipeline}.

Having constructed PSF models, we then proceed to measure the shapes of galaxies in the HST-ACS data. For this we make use of previous work by \cite{2010ApJ...708..202M} to serve as our source position catalogue. This analysis provides a photometric catalogue containing 438,226 sources with a source number density of $\sim74$ arcmin$^{-2}$ in the 2 deg$^{2}$ field, far surpassing the radio source density of $0.14$ arcmin$^{-2}$. The catalogue was extracted using \textsc{SExtractor}; for a detailed description see \cite{2010ApJ...708..202M}.

\begin{figure}
 \centering
 \includegraphics[width=0.475\textwidth]{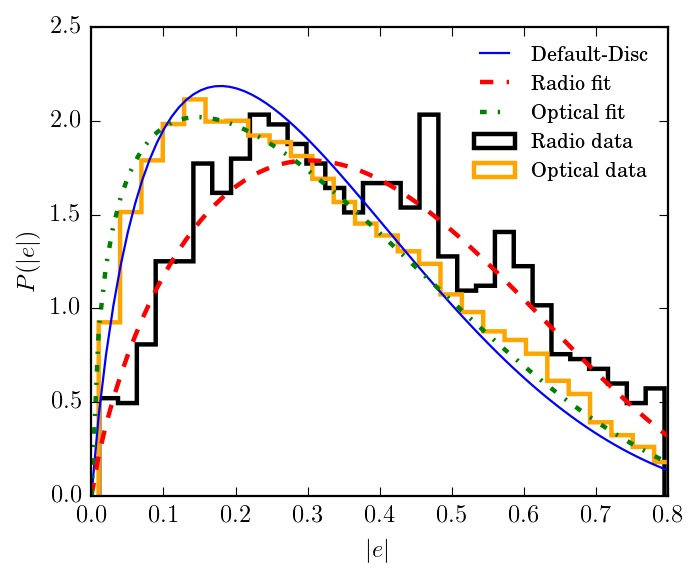}
 \caption[Radio $|\boldsymbol{e}|$ distribution]{Galaxy ellipticity distributions measured from the radio and optical COSMOS data. An ellipticity distribution model appropriate for disk galaxies (\cref{eq:STEP}) is shown in blue, for model parameters $B = 0.19$ and $C = 0.58$. The VLA COSMOS galaxy $|\boldsymbol{e}|$ measurements are shown as the black histogram which is best-fitted by \cref{eq:STEP} using B and C values of $0.113\pm 0.041$ and $0.303\pm 0.058$ respectively (shown as the red dotted line). The ellipticity distribution as measured from the HST optical data (shown as the orange histogram) is best-fit with model parameters $B = 4.65\pm{0.288\E^{-5}}$ and $C = 0.218\pm0.002$ (shown as the green dash-dot line).}
 \label{fig:E_hist_opt}
\end{figure}

For each galaxy in this catalogue, we measure the optical source shape with \textsc{im3shape}, making use of the PSF models described above. The resulting optical shape catalogue can be found at \url{https://github.com/bentunbridge/cosmos-radio-optical}. Once again, we refer the reader to \cref{fig:E_hist_opt} which compares the ellipticity distributions as measured in the optical and in the radio. For the optical data, we measure $\sigma_{e}$ = $0.26$ per ellipticity component with a model fit of $B = 4.65\pm{0.288\E^{-5}}$ and $C = 0.218\pm0.002$. The fitted parameters for both radio and optical ellipticity distributions are summarised in \cref{Tab:STEP}.

\begin{table}
\centering
\begin{tabular}{lccc}
\hline
Study & $\sigma_{e}$ & $B$ & $C$ \\ 
\hline
 STEP &$0.3$&$0.19$&$0.58$\\
 Optical &$0.26$&$4.65\pm{0.288\E^{-5}}$&$0.218\pm0.002$\\
 Radio &$0.29$&$0.137\pm 0.024$&$0.318\pm 0.032$\\
\hline
\end{tabular}
\caption[Optical and Radio $|\boldsymbol{e}|$ distribution]{Comparison of the measured ellipticity dispersion and ellipticity distribution fit parameters for the radio and optical datasets. Also listed are the parameters adopted for the representative model used in the GREAT08 analysis of \cite{2010MNRAS.405.2044B}.}\label{Tab:STEP}
\end{table}

\subsection{Radio-optical shape correlations}
\begin{figure*}
 \centering
  \includegraphics[width=\textwidth]{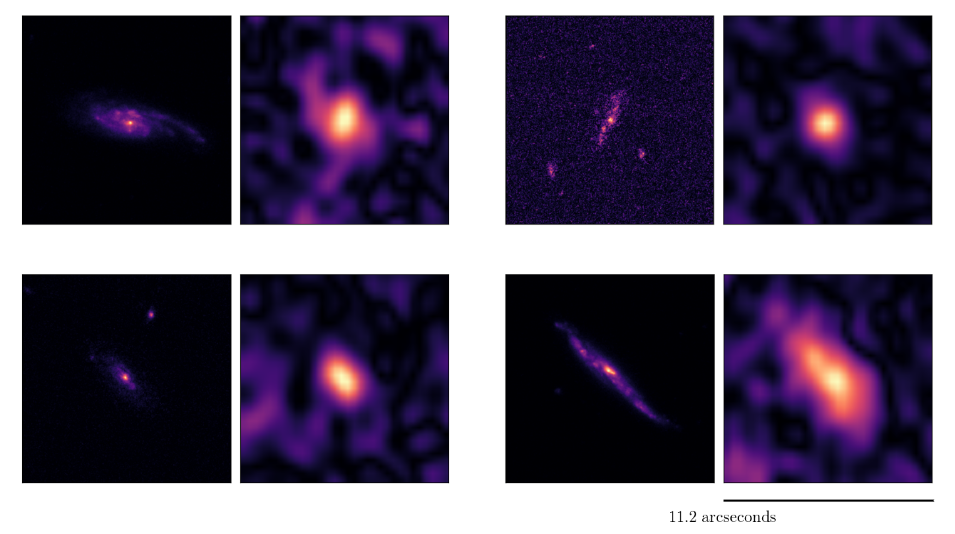}
  \caption[Optical - Radio source cutouts]{Comparison between the radio and optical images for 4 sources in the catalogue. In each case a $11.2''\times 11.2''$ cutout is shown for the optical (\emph{left panels}) and radio (\emph{right panels}) images of the matched source. Note the much poorer resolution of the radio images.}\label{fig:Source_compare}
\end{figure*}

With possession of shape measurements in both the optical and radio observations, we perform a cross-wavelength comparison of the source position angles. 
Motivated by the simulation results of \cref{simulation_err} we perform the comparison with the position angle estimates, also noting the existence of shear estimators which make use of position angles \citep{2014MNRAS.445.1836W}. We make individual cut-out images comparing the source shapes in radio and optical available at \url{https://github.com/bentunbridge/cosmos-radio-optical} -- we show a selection of these in \cref{fig:Source_compare}.


To aid a statistical comparsion, we also compute the associated uncertainties in the position angle measurements, $\sigma_{\alpha}$, for both the radio and optical shape catalogues. For the radio catalogue, the scatter in the position angles estimated from the VLA-COSMOS simulations (see \cref{simulation_err}) are used to estimate $\sigma_{\alpha}$ as a function of $S/N$. For the optical measurements we take advantage of previous work to estimate the uncertainty in an image-plane ellipticity measurement for a source with a given $S/N$ in the case of un-correlated pixel noise. The statistical uncertainties were derived analytically by \cite{2012MNRAS.427.2711K} and \cite{2012MNRAS.425.1951R} who also demonstrated the validity of these expressions using simplified galaxy models including multiplicative and additive noise with the \textsc{im3shape} code. For ellipticity components this uncertainty estimate is given by,
\begin{equation}
    \sigma_{\mathit{e}_{i}} = \frac{(1-\mathit{e}_{i}^{2})}{S/N},\\
\end{equation}
where $S/N$ is the ideal estimate of the signal-to-noise of the galaxy. Note that for nearly circular sources, i.e. small $\mathit{e}$ value, the statistical error associated with shape measurements saturates. Hence $\sigma_{\alpha}$ is derived through propagation from \cref{eq:pa}.

\begin{figure}
 \centering
 \includegraphics[width=0.475\textwidth]{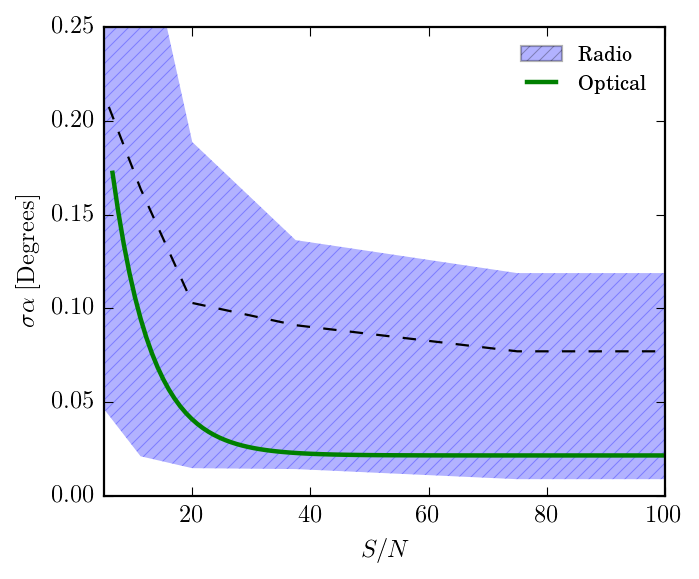}
 \caption[Radio and Optical $\sigma_{\alpha}$ as a function of $S/N$]{Average uncertainty in position angle measurements for both the radio and optical samples as a function of $S/N$. Radio uncertainties for the natural weighting scheme is derived from simulations and shown by the dashed line for our sample of radio galaxies. The hatched area shows this to a confidence level of $68\%$. The optical uncertainties are calculated analytically following \cite{2012MNRAS.425.1951R}.}
 \label{fig:SigmaPArecovery}
\end{figure}

\cref{fig:SigmaPArecovery} shows a comparison of the average uncertainty in the position angle ($\sigma_{\alpha}$) as a function of $S/N$ for both the radio and optical bands, where for the radio we have included the uncertainties for the naturally imaged weighting scheme, as used in this study. As can be seen, at low $S/N$ the radio position angles are estimated with a similar precision as is achieved in the optical. However the improvement seen in the $\alpha$ recovery at increased $S/N$ levels in the optical is not mirrored to the same extent in the radio data. 

Radio and optical sources are cross-matched to within a positional tolerance of $1.0$ arcsec and the position angles of the matched sample are cross-correlated. Within the uncertainties we do not make a statistically significant detection of correlation across the sample as a whole. We measure the Pearson`s correlation coefficient in $\alpha$ and find $R_{\alpha}=0.024$ and $R_{\alpha}=0.028$ for the images created using the natural weighting respectively. These low values of $R_{\alpha}$ indicate a clear lack of correlation between the data sets (a strong positive correlation would give $R_{\alpha}=1$).

\begin{figure}
 \centering
 \includegraphics[width=0.475\textwidth]{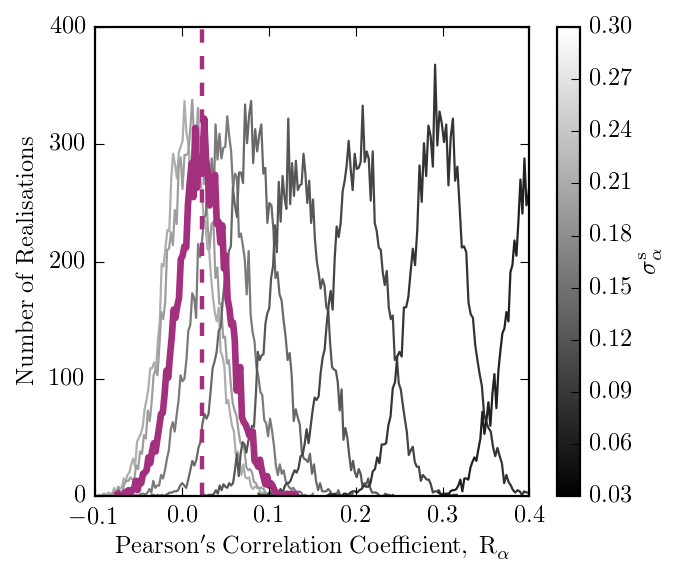}
 \caption[Radio Vs Optical Correlation Function Analysis]{Comparison of the distribution of Pearson correlation coefficients ($R_\alpha$) measured from simulations of matched catalogues for varying levels of intrinsic (astrophysical) scatter in the radio-optical position angle correlation as detailed by the colour scale ($\sigma^s_\alpha = \sigma^{\rm scatter}_\alpha / \pi$). The vertical dashed line shows the corresponding measurement from the data ($\approx 0.24$). For comparison we highlight the most consistent histogram, shown by the bold purple histogram.} 
 \label{fig:radioPArecovery}
\end{figure}

Although we do not detect a correlation in the radio and optical shapes, we investigate further in order to place limits on a global physical correlation. Assuming that optical and radio galaxy position angles do have some association, we may place limits on the level of astrophysical scatter ($\sigma_{\alpha}^{\mathrm{scatter}}$) required to produce the observed lack of correlation, taking into account our measurement errors. To do this, we conduct Monte Carlo (MC) simulations of catalogues of radio and optical matched objects. Taking into account the measurement errors, we incrementally increase the level of astrophysical scatter between the simulated radio and optical shapes, each time computing the $R_{\alpha}$ value for $10,000$ realisations. The results are shown in \cref{fig:radioPArecovery} where we also compare to the $R_\alpha$ value measured from the real data. We emphasise that our MC simulations fully capture both the shape measurement errors and the statistical error due to the finite number of sources in the data.

\begin{figure}
 \centering
 \includegraphics[width=0.475\textwidth]{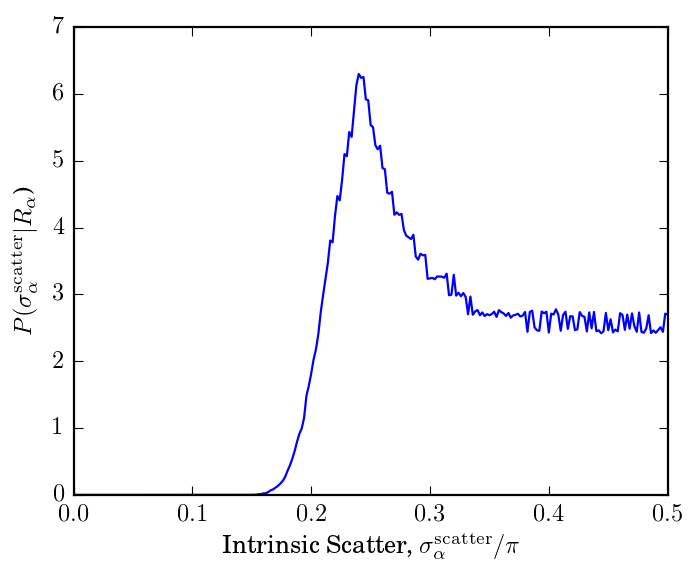}
 \caption[Radio Vs Optical, $P(\sigma_{\alpha})$ Analysis]{Limits on the intrinsic (astrophysical) scatter between the position angles of the same galaxies in the radio and optical bands. The curves show $P(\sigma_{\alpha}^{\mathrm{scatter}}|R_{\alpha})$ -- the probability distribution of $\sigma_{\alpha}^{\mathrm{scatter}}$ given the measured value of $R_\alpha$. The results obtained using the natural weighted radio image is shown as the blue curve.}
 \label{fig:sigma(Rpa)}
\end{figure}

The results are further summarised in \cref{fig:sigma(Rpa)} where we plot $P(\sigma_{\alpha}^{\mathrm{scatter}}|R_{\alpha})$: the probability distribution of $\sigma^{\rm scatter}_\alpha$ given the value of $R_{\alpha}$ measured from the data. The measured $R_{\alpha}$ is shown to be most consistent with a $\sigma_{\alpha}^{\mathrm{scatter}}$ value of $0.24\pi$ radians (or $14.3^{\circ}$) and the distribution corresponds to a $95\%$ lower limit of $0.212\pi$ radians (or $38.2^{\circ}$) on $\sigma_{\alpha}^{\mathrm{scatter}}$.

\subsection{Relation to earlier results}\label{Relation to Earlier Results}
Although the shapes of galaxies as measured in the optical and radio need not necessarily be aligned, previous studies have suggested such an alignment, at least for certain galaxy types. In particular, \cite{2009MNRAS.399.1888B} were able to show a strong alignment of the radio and optical images of galaxies classified as late-type and a weaker \emph{anti}-alignment correlation in position angle for sources classified as early-type galaxies. A physical motivation for this pattern can be made from the source of radio emission in each case. For late-type galaxies, radio emission is related to areas of recent star formation and so is expected to trace the similarly generated optical emission. In early-type galaxies the radio emission is driven by accretion on to a central black hole, producing bright jets anti-aligned with the galaxy disc, causing radio emission to be aligned with the optical minor axis. \cite{2009MNRAS.399.1888B} made use of the Sloan Digital Sky Survey (SDSS) and the Faint Images of the Radio Sky at Twenty centimetres (FIRST) for the optical and radio observations respectively. The key difference with the work presented here is the comparatively large volumes probed by SDSS and FIRST, compared to COSMOS. Between SDSS and FIRST $239,993$ matches are made between the optical and radio observations (although this is before the application of any selection cuts), far surpassing the $761$ sources used in this study.

In order to make a comparison between the studies, we implemented similar selection cuts to those used by \cite{2009MNRAS.399.1888B}. Their cuts included only considering bright sources via a Magnitude cut-off $M>18$, and a limitation on the ellipticity, implemented as $b/a <0.8$ (where $a$ and $b$ are the semi-major and semi-minor axis respectively), for both the optical (SDSS) and radio (FIRST) sample. Finally, a cut on the minimum radio source size was applied ($a>2^{\prime\prime}$) resulting in a sample of $14302$ galaxies. We modify these cuts due to the differences in the observation properties. In particular, we choose not to implement a magnitude cut since we have already applied $S/N$ cuts ($S/N>10$ and $S/N>5$ for the optical and radio samples respectively). We impose the same cuts on ellipticity, only including sources with $b/a <0.8$ in both the optical (HST-ACS) and radio (VLA) samples. We impose a less stringent size cut ($a>1^{\prime\prime}$) compared to that used in \cite{2009MNRAS.399.1888B} due to the increased resolution of the COSMOS data.\footnote{The FIRST data used in \cite{2009MNRAS.399.1888B} has a resolution of $5^{\prime\prime}$, compared to $\sim1.5^{\prime\prime}$ for the VLA-COSMOS data.}

After implemented these cuts, only $327$ sources remain. Of these $327$, we can further split the sample into galaxy morphology types by matching to a previous COSMOS morphology catalogue compiled by \cite{2006astro.ph.11644S}. The results are summarised in \cref{Tab:class}. Correlating the radio and optical shapes within these classification subsets, we again find no meaningful correlation between the radio and optical shapes. We note however that the number of galaxies in our sub-samples are small. We further note that the main sample of galaxies used in this study (being composed mainly of faint, high-redshift star-forming galaxies) will be a much closer match to a typical weak lensing galaxy sample, compared to the sample identified in \cite{2009MNRAS.399.1888B}.

\begin{table}
\centering
\begin{tabular}{lcc}
\hline
Type & Number & Fraction \\
\midrule
 Early & 59 & 0.18 \\
 Late & 120 & 0.37 \\
 Irregular & 122 & 0.37 \\
 No Classification & 26 & 0.08 \\
\hline
\end{tabular}
\caption[Optical to Radio Match, Sample Size with Strict Cuts]{Sub-samples of the COSMOS radio-optical matched catalogue, after undergoing further selection cuts, and sub-divided according to galaxy type. From the full sample of radio galaxies detected in our VLA-COSMOS image we find $761$ galaxies are successfully matched to an optical source. However only $327$ remain after the applying the cuts described in  \cref{Relation to Earlier Results}.}\label{Tab:class}
\end{table}

An apparent lack of correlation between optical and radio galaxy shapes has been commented on before by \cite{2010MNRAS.401.2572P}. This study attempted to measure a weak lensing signal in the radio with eMERLIN and VLA observations, while simultaneously conducting a cross correlations study of galaxy shapes. Subsequently they found no correlation between the optical and radio shapes, which is consistent with our findings here.


\section{Conclusions}\label{concl}
In this paper we have performed a detailed comparison analysis of the shapes of galaxies in the COSMOS field, as measured in the optical using HST, and as measured in the radio, using the VLA. Our study has been motivated by the scientific potential of cross-correlation cosmic shear analyses of future overlapping optical and radio surveys \citep{2015aska.confE..23B, 2016arXiv160103947H, 2016arXiv160103948B, 2016arXiv160603451C}. In order to fully exploit such future cross-correlations, one needs to understand in detail the correlations in intrinsic optical and radio shapes, which is the issue that we have attempted to address in this study.

In the course of our analysis, we have highlighted some of the challenges involved in extracting shapes from radio observations. For this analysis we have chosen to measure galaxy shapes from images reconstructed from the VLA radio data. In particular, we have used simulations, composed of a known distribution of galaxy shapes combined with a realistic radio interferometer observation and imaging pipeline, to show the effectiveness of position angle ($\alpha$) recovery through typical radio data reduction and image creation techniques. We have investigated the use of two variations of the widely used H\"ogbom-\clean image creation algorithm in the radio: natural and uniform visibility weighting. We find the choice in weighting scheme affects the measured shape parameters greatly. In particular the choice of a uniform weighting scheme yields largely discrepant values of multiplicative bias in ellipticity components for a given galaxy. We found this problem was diminished when natural weighting was used, which produced highly similar multiplicative bias on ellipticities, which later cancels when position angles are derived from the measured ellipticities. This allowed us to obtain unbiased estimates of the position angles, which were important for this study.

Quantifying the bias in position angle ($\alpha$) achieved in this study with a linear bias model, a link to weak lensing requirements for future surveys was established via the position angle only shear estimators of \cite{2014MNRAS.445.1836W}. A key result is shown in \cref{fig:radioarecovery} which allows one to translate from the usually quoted requirements on shear bias to requirements on position-angle bias. We find that although our position angle recovery appears relatively unbiased given our measurement errors, the uncertainty on the position angle bias that we are able to achieve using current data is still larger than current weak lensing requirements.

After quantifying the expected uncertainties through simulations we have applied our shape measurement pipeline to the real COSMOS radio (VLA) observations. We used the resulting radio galaxy shape measurements (and associated uncertainties) to place a lower limit on the astrophysical scatter in source position angles between the continuum radio and optical emission of the sources which are detected in both our VLA data and an optical HST study of the same field. We find this lower limit to be $\sigma_{\alpha}^{\mathrm{scatter}} \gtrsim 0.212\pi$ (or $38.2^{\circ}$) at a $95\%$ confidence level. This appears consistent with results from previous studies \citep[][]{2010MNRAS.401.2572P} considering the low absolute number of sources in our sample. 

Understanding the radio-optical shape correlations is important as it will affect the noise term on the cosmological power spectra measured by radio-optical cross-correlation weak lensing studies \citep{2016arXiv160103947H, 2016arXiv160603451C}. High levels of correlation will increase the shot noise on shear measurement, but allow for cross-waveband calibration against some systematics, whilst low levels of correlation effectively increase the source number density being used to probe the cosmic shear field. In the near future deep, high-resolution optical and radio surveys such as SuperCLASS\footnote{\url{http://www.e-merlin.ac.uk/legacy/projects/superclass.html}}
will further constrain this correlation, better informing forecasts and survey design considerations for SKA. 

\section*{Acknowledgements}
We thank Chris Hales, Anita Richards, Neal Jackson, Lee Whittaker and Joe Zuntz for useful discussions. We further thank Chris Hales, Eva Schinnerer and Vernesa Smolcic for providing us with the partially processed VLA data. MLB is an STFC Advanced/Halliday fellow. BT, IH and MLB are supported by an ERC Starting Grant (grant no. 280127).

\label{Bibliography}
\bibliographystyle{mn2e_plus_arxiv}
\bibliography{radio_shape_COSMOS}{}

\begin{appendices}
\crefalias{section}{appsec}
\section{Position angle to shear bias}\label{shear_bias_from_alpha_bias}
In this section we relate our shape recovery in terms of shear. It is important to propagate obtained bias in position angle recovery (see \cref{simulation_err}) to the actual quantity of interest for scientific inference: the measured shear which can be related to the line of sight gravitational potential. Requirements on shear measurement additive and multiplicative systematics for these errors to be smaller than statistical uncertainties \citep[calculated according to the prescription derived in][]{2008MNRAS.391..228A} are given in \cref{Tab:m_and_c_shear}.
\begin{table}[h]
\centering
\begin{tabular}{lcc}
\hline
 Survey&  $|m|<$ & $|c|<$ \\
\hline
Stage II & $0.02$ & $0.001$ \\
Stage III & $0.004$ & $0.0006$ \\
Stage IV & $0.001$& $0.0003$ \\
\hline
\end{tabular}
\caption[$m$ and $c$ for Weak Lensing Surveys]{Requirements on the systematic bias on shear components ($\gamma_{1}$ and $\gamma_{2}$), expressed in terms of the multiplicative ($m$) and additive ($c$) components of a linear bias model. Requirements are given for completed surveys (``Stage II" such as CFHTLens), intermediate-sized ongoing or future surveys (``Stage III" such as DES and SKA1) and for high-precision future surveys (``Stage IV" such as \emph{Euclid} and SKA2).}\label{Tab:m_and_c_shear}
\end{table}

For shear estimation from ellipticities using the na\"{i}ve estimator,
\begin{equation}
\label{eqn:gamma_hat}
\hat{\gamma} = \frac{1}{N}\sum_{i=1}^{N}e^{\mathrm{obs}}_{i},
\end{equation}
these requirements on shear translate readily to requirements on ellipticity.  However for this study we are interested in exploiting our relatively unbiased measurements of the position angles $\alpha$. We therefore need to understand how biases in position angle measurements propagate into biases in the derived shear. 

Position angle-only shear estimators have recently been highlighted in \cite{2014MNRAS.445.1836W} and shown to be competitive with other current weak lensing estimators using the \textsc{great3} simulations in \cite{2015MNRAS.454.2154W}. We follow this work and apply the third order $\alpha$-only shear estimator of \cite{2014MNRAS.445.1836W} to multiple realisations of a sample of galaxies. This shear estimator is given by \citep{2014MNRAS.445.1836W}
\begin{align}\label{eq:3rdorder}
\hat\gamma &= \sqrt{\frac{32\sigma_{\mathit{e}}^{2}}{2\hat F_{1}^{2} N^{2}}}   \times
\\ \nonumber
& \cos \left [\frac{1}{3}\left(\tan^{-1} \left( \sqrt{\frac{4 \pi}{27 \mathit{e}^{4\sigma_{\alpha}^{2}} \hat F_{1}^{2}} -1 }\right) + \pi \right)\right ] \sum^{N}_{i=1} \boldsymbol{n(i)},
\end{align}
where the vector $\boldsymbol{n(i)}$ is a 2-element vector composed of the sine and cosine of $2\alpha$ for the $i^{th}$ galaxy, $\sigma_{\mathit{e}}$ is the intrinsic dispersion in galaxy ellipticities and $N$ is the total number of galaxies in the sample. $F_{1}(|\hat\gamma|)$ is a function given by,
\begin{align}\label{eq:F1}
F_{1}(|\hat\gamma|) &= \nonumber \\
& \mathit{e}^{2\sigma_{\alpha}^{2}}   \sqrt{\left[\frac{1}{N} \sum^{N}_{i=1} \cos(2\alpha^{(i)}) \right]^{2} + \left[\frac{1}{N} \sum^{N}_{i=1} \sin(2\alpha^{(i)}) \right ]^{2}},
\end{align}
where $\sigma_{\alpha}$ is simply the intrinsic dispersion of measurement errors on the observed (sheared) position angles.
The third order approximation (which requires inversion and interpolation of the $F_{1}(|\hat\gamma|)$ function) is a good approximation to the full estimator of \cite{2014MNRAS.445.1836W} in the regime $|\boldsymbol{\gamma}| \lesssim 0.2$.

In order to relate the accuracy with which we can recover position angles (as demonstrated in \cref{fig:radioalpharecovery}) to the requirements on shear recovery for representative weak lensing experiments (as listed in \cref{Tab:m_and_c_shear}), we have applied the third-order position-angle only shear estimator to a set of simple simulations that include linear shear biases in the simulated position angles.

We simulate position angle measurements with varying levels of intrinsic linear bias in $\alpha$, in the range of $-0.2<m_{\alpha}<0.2$ and $-0.2<c_{\alpha}<0.2$. This was sampled using a gridded mesh of $200\times800$ upon the $m_{\alpha}$ and $c_{\alpha}$ parameter space to select input values, totalling $160,000$ realisations. For each realisation value we simulate a shear catalogue of $10,000$ galaxy shapes, each with a known gravitational shear component added to the galaxy shapes (where $\mathit{e}_{i}^{\mathrm{sky}} = \mathit{e}_{i}^{\mathrm{int}} + \gamma_{i}$). We then calculate the true position angles of the simulated galaxies ($\alpha_i^{\rm sky}$) using \cref{eq:posn_angle_def} and add a position-angle measurement bias according to:
\begin{equation}\label{eq:alpha_obs}
\alpha_{i}^{\mathrm{obs}}= (m_{\alpha} + 1) \alpha_{i}^{\mathrm{sky}}  + c_{\alpha},
\end{equation}
where $\alpha_{i}^{\rm obs}$ is now the ``observed" galaxy position angle (including both shear and bias effects). We then apply the third order position-angle only shear estimator described above to estimate the shear from our simulated position angle measurements. We repeat this process for several input shear values in order to fit a linear bias model for the estimated shear. The resultant shear bias parameters can then be compared to the shear bias requirements of \cref{Tab:m_and_c_shear} in order to place corresponding requirements on the bias in position angle measurements, i.e. on $m_\alpha$ and $c_\alpha$. \cref{fig:radioarecovery} in \cref{Position angle measurement bias} shows the bias in the estimated shear that results from the position angle bias for a range of position angle bias parameters. The performance of the position-angle recovery from the VLA-COSMOS simulations (see \cref{sim_bias_estimation}) and the requirements of \cref{Tab:m_and_c_shear} are also indicated on this figure.

\end{appendices}

\end{document}